\newif\ifsingle\singletrue

\newif\ifproofs\proofstrue 

\ifsingle
\documentclass[12pt,draftclsnofoot, onecolumn]{IEEEtran}		
\else		
\documentclass[9pt,final, twocolumn]{IEEEtran}
\fi

\usepackage{times}
\usepackage{amsmath,dsfont}
\usepackage{amssymb,amsthm}
\usepackage{epsfig,verbatim}
\usepackage{subfigure}
\usepackage{setspace}
\usepackage{color}
\usepackage{cite}
\usepackage{epstopdf}
\usepackage{graphics}
\usepackage{accents}
\usepackage{acronym}
\usepackage{booktabs}
\usepackage{mathtools}
\usepackage{enumitem}

\usepackage[ruled,linesnumbered]{algorithm2e}  
\SetKwInput{KwData}{\textbf{Init}} 

\newcommand{\myVec}[1]{{\boldsymbol{#1}}}
\newcommand{\myMat}[1]{{\boldsymbol{#1}}}
\newcommand{\mySet}[1]{\mathcal{#1}}
\newcommand{\E}{\mathds{E}}		 			

\newtheorem{proposition}{Proposition}
\newtheorem{lemma}{Lemma}

    \def\Complex{{\rm\rule[.23ex]{.03em}{1.1ex}\kern-.3em{C}}}

    \newcommand{\be}{\begin{equation}} \newcommand{\ee}{\end{equation}}
    \newcommand{\bea}{\begin{eqnarray}} \newcommand{\eea}{\end{eqnarray}}
    \newcommand{\benum}{\begin{enumerate}} \newcommand{\eenum}{\end{enumerate}}

    \newcommand{\qq}{{\bf q}}
    
    \newcommand{\qs}{{\bf s}}

    \newcommand{\qw}{{\bf w}}
    
    \newcommand{\qy}{{\bf y}}
    \newcommand{\qz}{{\bf z}}

    \newcommand{\qA}{{\bf A}}
    \newcommand{\qB}{{\bf B}}
    \newcommand{\qC}{{\bf C}}
    
    \newcommand{\qE}{{\bf E}}
    \newcommand{\qF}{{\bf F}}
    \newcommand{\qG}{{\bf G}}
    \newcommand{\qH}{{\bf H}}
    \newcommand{\qI}{{\bf I}}
    
    \newcommand{\qK}{{\bf K}}

    \newcommand{\qN}{{\bf N}}
    \newcommand{\qO}{{\bf O}}
    \newcommand{\qP}{{\bf P}}
    \newcommand{\qQ}{{\bf Q}}
    
    \newcommand{\qS}{{\bf S}}
    
    \newcommand{\qU}{{\bf U}}
    \newcommand{\qV}{{\bf V}}
    \newcommand{\qW}{{\bf W}}

    \newcommand{\qZ}{{\bf Z}}

    \newcommand{\qPsi}{{\boldsymbol \Psi}}
    \newcommand{\qPhi}{{\boldsymbol \Phi}}
    \newcommand{\qXi}{{\boldsymbol \Xi}}

    \newcommand{\qSigma}{{\boldsymbol \Sigma}}
    
    \newcommand{\qUpsilon}{{\boldsymbol \Upsilon}}

    \newcommand{\calC}{{\cal C}}
    \newcommand{\calD}{{\cal D}}

    \newcommand{\calM}{{\cal M}}
    \newcommand{\calN}{{\cal N}}

    \newcommand{\calQ}{{\cal Q}}

    \newcommand{\tr}{{\sf tr}}

\ifsingle
\newcommand{\figWidth}{0.65\columnwidth}

\else
\newcommand{\figWidth}{0.75\columnwidth}

\fi 

\ifproofs
\newcommand{\myFootnote}[1]{\footnote{#1}}
\else
\newcommand{\myFootnote}[1]{ }
\fi

\acrodef{adc}[ADC]{analog-to-digital convertor}  
\acrodef{csi}[CSI]{channel state information} 
\acrodef{snr}[SNR]{signal-to-noise ratio}
\acrodef{bs}[BS]{base station}  
\acrodef{ber}[BER]{bit error rate}  
\acrodef{mu}[MU]{multi-user}
\acrodef{mimo}[MIMO]{multiple-input multiple-output}
\acrodef{mse}[MSE]{mean-squared error}
\acrodef{mmse}[MMSE]{minimal \ac{mse}}
\acrodef{pdf}[PDF]{probability density function}
\acrodef{rv}[RV]{random variable} 
\acrodef{isi}[ISI]{intersymbol interference}  
\acrodef{awgn}[AWGN]{additive white Gaussian noise} 
\acrodef{ut}[UT]{user terminal} 
\acrodef{mmw}[mmWave]{millimeter wave}
\acrodef{dma}[DMA]{dynamic metasurface antenna}
\acrodef{ofdm}[OFDM]{orthogonal frequency division multiplexing}
\acrodef{cp}[CP]{cyclic prefix} 
\acrodef{dft}[DFT]{discrete Fourier transform}
\acrodef{dtft}[DTFT]{discrete-time Fourier transform}
\acrodef{rf}[RF]{radio frequency}

\title{Dynamic Metasurface Antennas for MIMO-OFDM Receivers with \\Bit-Limited ADCs}
\author{
	\IEEEauthorblockN{Hanqing Wang, Nir Shlezinger, Yonina C. Eldar, Shi Jin, \\ Mohammadreza F. Imani, Insang Yoo, and David R. Smith
	} 
	\thanks{This project has received funding from the Benoziyo Endowment Fund for the Advancement of Science, the	Estate of Olga Klein -- Astrachan, the European Union’s Horizon 2020 research and innovation program under grant No. 646804-ERC-COG-BNYQ, and the Air Force Office of Scientific Research under grants No. FA9550-18-1-0187 and FA9550-18-1-0208.
		H. Wang and S. Jin are with the National Mobile Communications Research Lab, Southeast University, Nanjing,  P. R. China (e-mail: 
		\{hqwanglyt; jinshi\}@seu.edu.cn). 	
		N. Shlezinger  and Y. C. Eldar are with the Faculty of Math and CS, Weizmann Institute of Science, Rehovot, Israel (e-mail: \{nir.shlezinger; yonina\}@weizmann.ac.il). 	
		M. F. Imani, I. Yoo, and D. R. Smith are with the Department of ECE, Duke University, Durham, NC (e-mail:  mohamad.imani@gmail.com; \{insang.yoo, drsmith\}@duke.edu).
	}

	\vspace{-1.0cm}
	
}
\vspace{-0.75cm}

\begin{document}
	
	\maketitle
	\begin{abstract}
		The combination of orthogonal frequency modulation (OFDM) and multiple-input multiple-output (MIMO) systems plays an important role in modern communication systems. In order to meet the growing throughput demands, future MIMO-OFDM receivers are expected to utilize a massive number of antennas, operate in dynamic environments, and explore high frequency bands, while satisfying strict constraints in terms of cost, power, and size. An emerging technology to realize  massive MIMO receivers of reduced cost and power consumption is based on dynamic metasurface antennas (DMAs), which inherently implement controllable compression in acquisition.  
		In this work we study the application of DMAs for MIMO-OFDM receivers operating with bit-constrained analog-to-digital converters (ADCs). 
		We present a model for DMAs which accounts for the configurable frequency selective profile of its metamaterial elements, resulting in a spectrally flexible hybrid structure.
		We then exploit previous results in task-based quantization to show how DMAs can be configured to improve recovery in the presence of  constrained ADCs, and propose methods for adjusting the DMA parameters based on channel state information. Our numerical results demonstrate that the DMA-based receiver is capable of accurately recovering OFDM signals. In particular, we show that by properly exploiting the spectral diversity of DMAs, notable performance gains are obtained over existing designs of conventional hybrid architectures, demonstrating the potential of DMAs for MIMO-OFDM setups in realizing high performance massive antenna arrays of reduced cost and power consumption.

		{\textbf{\textit{Index terms---}} Metasurface antennas, quantization, MIMO-OFDM.}
	\end{abstract}
	
	
	\section{Introduction}
	Wireless networks are subject to constantly growing throughput demands. To satisfy these requirements, cellular \acp{bs} are equipped with a large number of antennas while serving multiple remote users \cite{Marzetta-2010TWC}, utilizing wideband \ac{ofdm} transmissions. Such \ac{mu} \ac{mimo} \ac{ofdm} architectures are capable of reliably providing increased data rates to a large amount of users~\cite{jiang2007multiuser}.
	
	In addition to their performance requirements, \acp{bs} are expected to be cost  efficient, operate under strict power constraints, and support deployment in various physical shapes and sizes. 	A major challenge associated with realizing such \ac{mu}-\ac{mimo}-\ac{ofdm} systems stems from  the increased cost of \acp{adc} \cite{Eldar2015}, which allow the analog signals observed by each antenna  to be processed in digital. The power usage of an \ac{adc} is  related to the signal bandwidth and the number of bits used for digital representation \cite{walden1999analog}. Thus, when the number of antennas and \acp{adc} operating at wide bands is large, limiting the number of bits, thus operating under quantization constraints, is crucial to keep feasible cost and power usage~\cite{Andrews-2014JSAC}.  
	
	Focusing on uplink communications, quantization constraints imply that the \ac{bs} cannot process the channel output directly but rather  a discretized distorted representation of it. The distortion induced by the continuous-to-discrete quantization mapping   degrades the ability to  extract the desired information, such as recovering the transmitted signal, from the observed channel output. 
	An attractive strategy to mitigate the effect of quantization error is to incorporate pre-quantization processing in analog resulting in a hybrid architecture. Jointly designing the analog processing along with the quantization rule and the digital mapping, referred to as {\em task-based quantization}, was shown to facilitate recovery of the underlying information in the digital domain \cite{shlezinger2018hardware,salamatian2019task,shlezinger2019deep}. An advantage of such  hybrid \ac{mimo} receivers, originally proposed as a method to decrease the number of \ac{rf}  chains \cite{zhang2005variable,mendez2016hybrid,ioushua2019family,sohrabi2017hybrid}, is that they can be used to reduce the number of quantized samples, and accordingly, the number of \acp{adc},  compared to assigning an \ac{adc} to each antenna \cite{shlezinger2018asymptotic}. Nonetheless, such designs require an additional dedicated hardware \cite{gong2019rf}, and  the  pre-quantization mapping, as well as the ability to adapt its parameters based on the channel conditions, is typically limited and dictated by the analog components~\cite{mendez2016hybrid, ioushua2019family}.
	
	An alternative receiver architecture which implements adjustable analog combining in the hardware level is based on \acp{dma} \cite{DSmith-2017PRA,smith2017analysis}. Such surfaces consist of a set of microstrips, each embedded with configurable radiating metamaterial elements \cite{ Sleasman-2016JAWPL,Diebold-2018AO}. 
	Recent years have witnessed a growing interest in the application of metasurfaces as reflecting surfaces for wireless communications \cite{huang2019reconfigurable,di2019smart, del2019optimally,tang2019wireless,dai2019wireless}. In such applications, a metasurface is placed in a physical location where it can aid the \ac{bs} by reflecting and steering the transmitted waveforms. 
	Metasurfaces utilized as antennas in wireless communications, i.e., as transmitting and receiving devices rather than configurable reflectors, were recently studied in \cite{shlezinger2019dynamic,wang2019dynamic,DSmith-2018TCOM}. Such antenna structures typically use much less power and cost less than architectures based on standard arrays \cite{Johnson-2016TAP}, while facilitating the implementation of a large number of tunable elements in a given physical area \cite{akyildiz2016realizing}. 
	In the context of wireless communications, it is shown in  \cite{shlezinger2019dynamic,wang2019dynamic}  that the achievable rate when utilizing \acp{dma} without quantization constraints is comparable to using ideal antenna arrays. The potential of \acp{dma} for realizing  massive \ac{mimo} antennas combined with the need of such systems to operate with low resolution quantization motivates the study of bit-constrained \ac{dma}-based \acp{bs}, which is the focus of the current work.
	
	Here, we study uplink \ac{mu}-\ac{mimo}-\ac{ofdm} communications in which a bit-constrained \ac{bs} is equipped with a \ac{dma}. 
	We first extend the model formulated in \cite{shlezinger2019dynamic}, which was built upon approximations of  the \ac{dma} properties proposed in \cite{DSmith-2017PRA} that hold for narrowband signals, to faithfully capture the reconfigurable frequency selective nature of \acp{dma} in wideband setups, such as \ac{ofdm} systems. Then, we  show how the resulting \ac{dma} characteristics can be incorporated into the \ac{mu}-\ac{mimo}-\ac{ofdm} model, resulting in a form of a hybrid receiver. However, while conventional hybrid architectures require a dedicated analog combining hardware whose mapping is typically frequency flat \cite{zhang2005variable,mendez2016hybrid,ioushua2019family, gong2019rf, roth2018comparison,mo2017channel,jacobsson2017throughput,choi2016near}, \acp{dma} implement a controllable frequency selective profile as an inherent byproduct of their antenna structure. We use this model to formulate the following problem: How can the dynamic properties of \acp{dma}, i.e., their configurable reception parameters, be exploited to facilitate the task of recovering the transmitted \ac{ofdm} signals from the output of low-resolution \acp{adc}?
	
	Based on this formulation, we cast the problem as a task-based quantization setup, in which the task  is to accurately recover the transmitted \ac{ofdm} symbols.  
	Using this framework, we derive a scheme  for jointly optimizing the \ac{dma} weights along with the quantization system, i.e., the \ac{adc} support and the digital processing, under a given bit constraint. Our proposed method consists of three algorithms.
	The first algorithm utilizes a greedy optimization method to tune the \ac{dma} weights, ignoring the structure constraints induced by the physics of these metasurfaces. The next two algorithms then identify a feasible approximation of the unconstrained \ac{dma} obtained using the first algorithm, where each technique applies to a different \ac{dma} model: 
	The first of these two methods is based on the approximations of \ac{dma} characteristics proposed in \cite{DSmith-2017PRA}, which ignore the spectral flexibility of the elements, tuning a frequency flat hybrid receiver most suitable for narrowband signals; The latter makes usage of the full frequency selective profile of the metamaterial elements, and is thus preferable for wideband transmissions.
	Both proposed algorithms exploit the unique structure of the hybrid system which arises from the task-based quantization framework, while building upon the dynamic nature of \acp{dma}, which allows to set their parameters in run-time in light of the channel conditions. 
	
	The performance of the resulting receivers designed using these proposed algorithms, in terms of 
	\ac{ofdm} signal recovery accuracy and uncoded \ac{ber}, are evaluated in a simulation study. Our numerical results demonstrate the ability of bit-constrained \acp{dma} to achieve notable performance gains over conventional hybrid architectures designed using previously proposed methods.  
	These performance gains, which arise from the combination of task-based quantization tools and the spectral flexibility of \acp{dma}, add to their practical benefits over conventional hybrid structures, which follow from the fact that \acp{dma} do not require additional dedicated hardware for implementing their analog processing.
	
	The rest of this paper is organized as follows: 
	Section~\ref{sec:Model}  presents the system model of \ac{dma}-based receivers in bit-constrained \ac{mimo}-\ac{ofdm} systems, introducing the formulation of the configurable frequency selectivity of the metamaterial elements and how it is incorporated in the wireless communication setup.
	Section~\ref{sec:DMA} details the proposed \ac{dma} design methods along with a discussion. 
	Numerical examples are presented in Section~\ref{sec:Sims}. 
	Finally,  Section~\ref{sec:Conclusions}  concludes the paper.
	Proofs of the  results stated in the paper are detailed in the appendix. 
	
	Throughout the paper, we use boldface lower-case letters for vectors, e.g., ${\myVec{x}}$;
	Matrices are denoted with boldface upper-case letters,  e.g., 
	$\myMat{M}$; and  
	calligraphic letters are used for sets, where $\mySet{C}$ and $\mySet{Z}$ are the complex numbers and integers, respectively. 
	The $\ell_2$ norm, Kronecker product, transpose, Hermitian transpose, conjugate, and trace are denoted by $\| \cdot \|$, $\otimes$,    $(\cdot)^T$, $(\cdot)^H$, $(\cdot)^*$, and ${\rm tr}[\cdot]$, respectively. We use $\qI_n$ and $\qO_n$ for the $n\times n$ identity matrix and all-zero matrix, respectively.
	Finally, ${\rm blkdiag}(\qA_1,\dots,\qA_n)$ is a block diagonal matrix with the matrices $\qA_1,\dots,\qA_n$ along its diagonal.

	\section{System Model}
	\label{sec:Model}
	In this section, we present the mathematical model for \ac{dma}-based uplink \ac{mu}-\ac{mimo}-\ac{ofdm} systems with bit-constrained \acp{adc}. We begin with the  model for the \ac{dma} operation in Subection \ref{subsec:DMA_model}, which extends the one proposed in our previous work \cite{shlezinger2019dynamic} to fully capture the frequency selective characteristics of metasurfaces used as antennas. 
	Then, in Subsection \ref{subsec:MIMO_OFDM}, we formulate the input-output relationship of  coarsely quantized uplink \ac{mu}-\ac{mimo}-\ac{ofdm} systems  in which the \ac{bs} is equipped with a \ac{dma}.
	
	
	\subsection{Dynamic Metasurface Antennas}
	\label{subsec:DMA_model} 
	Metamaterials are a class of artificial materials whose physical properties, such as their permittivity and permeability, can be externally configured to achieve some desired electromagnetic properties    \cite{DRSmith-2004Science}. 
	Metamaterial elements stacked in surface configurations are referred to as  metasurfaces. These two-dimensional structures can be tuned element-wise, allowing the metasurface to carry out desired operations, such as radiation, reflection, beamforming, and reception of propagating waves \cite{Smith-2012APM}.
	In particular, metasurfaces can be utilized as antennas by incorporating such surfaces on top of a guiding structure.
	A simple and common metasurface antenna architecture is comprised of a set of microstrips, each consisting of a multitude of sub-wavelength, frequency-dependent resonant metamaterial radiating elements \cite{Hunt-2013Science}, whose radiation properties are dynamically adjustable.  
	A larger antenna array can be thus formed by increasing the number of microstrips, or alternatively, by tiling several such metasurface antennas together.

\begin{figure}
		\centering	
		\includegraphics[width=0.7\columnwidth]{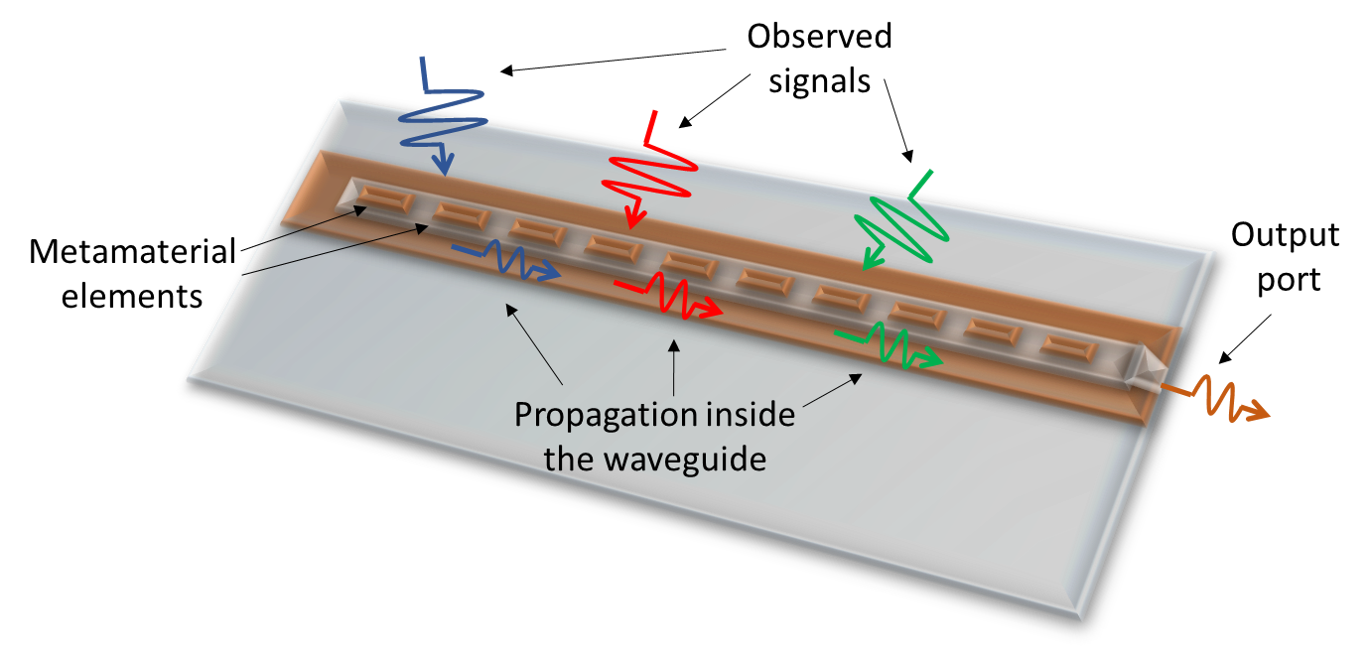}
		\vspace{-0.4cm}
		\caption{Illustration of signal reception using a microstrip.} 
		\label{fig:Microstrip}
	\end{figure}

	When used as a receive antenna, the signals observed by the elements are captured at a single output port for each microstrip, feeding an RF chain and an \ac{adc} with Nyquist rate sampling. An illustration of a set of observed signals captured using a single microstrip is depicted in Fig.~\ref{fig:Microstrip}. 
	The relationship between these signals and the micropstip output is dictated by the following two properties:
	\begin{enumerate}[label={\em P\arabic*}]
		\item \label{itm:P1} Each element acts as resonant electrical circuit, whose frequency response is described by the Lorentzian form \cite{DSmith-2017PRA, smith2017analysis}:
		\begin{equation}
		\label{eqn:FreqSel}
		\alpha(f) = \frac{F\cdot f^2}{(f^R)^2 - f^2 -j \chi f},
		\end{equation}
		where $F>0$ is the oscillator strength, $f^R>0$ is the resonance frequency, and $\chi>0$ is the damping factor. In \acp{dma}, these parameters can be varied by external control for each element individually \cite{Sleasman-2016JAWPL}.
		\item \label{itm:P2} Since the output port is located on the edge of the microstrip while the elements are uniformly placed along it, each signal which propagates from an element to that port undergoes a different path, and thus accumulates a different delay, depending on the specific element. In particular, letting $\beta$ be the wavenumber along the microstrip and $\rho_l$ denote the location of the $l$th element, this phase shift can be modeled as a filter whose frequency response is proportional to $e^{-j \beta \rho_l}$. 
	\end{enumerate}
	To facilitate the configuration of \acp{dma}, the frequency response of the metamaterial elements in \ref{itm:P1} is often assumed to be {\em frequency flat}, i.e., $\alpha(f) \equiv \alpha$ for each considered $f$, commonly representing narrowband signals. Under this assumption, the ability to externally control the response of the elements is typically modeled as allowing to set any $\alpha \in [a_{\min}, a_{\max}]$ for some $0 < a_{\min} < a_{\max}$, referred to as {\em amplitude-only weights} \cite[Sec. III-A]{DSmith-2017PRA}, or alternatively,  $\alpha \in \{\frac{j+e^{j \phi}}{2}| \phi \in [0,2\pi]\}$, referred to as {\em Lorentzian-constrained phase weights} \cite[Sec. III-D]{DSmith-2017PRA}. Our previous work \cite{shlezinger2019dynamic} used this frequency invariance assumption to analyze the achievable rate of uplink  \ac{dma}-based \ac{mimo} systems. Here, as we consider wideband \ac{ofdm} signals, we adopt the general model in \eqref{eqn:FreqSel} rather than its simplified narrowband approximation. In particular, by controlling the resonance frequency $f^R$ and the damping factor $\chi$, one can obtain a variety of different frequency selective profiles for each element. 
	
\begin{figure}
		\centering
		\begin{minipage}{0.45\textwidth}
			\centering
			\scalebox{0.48}{\includegraphics{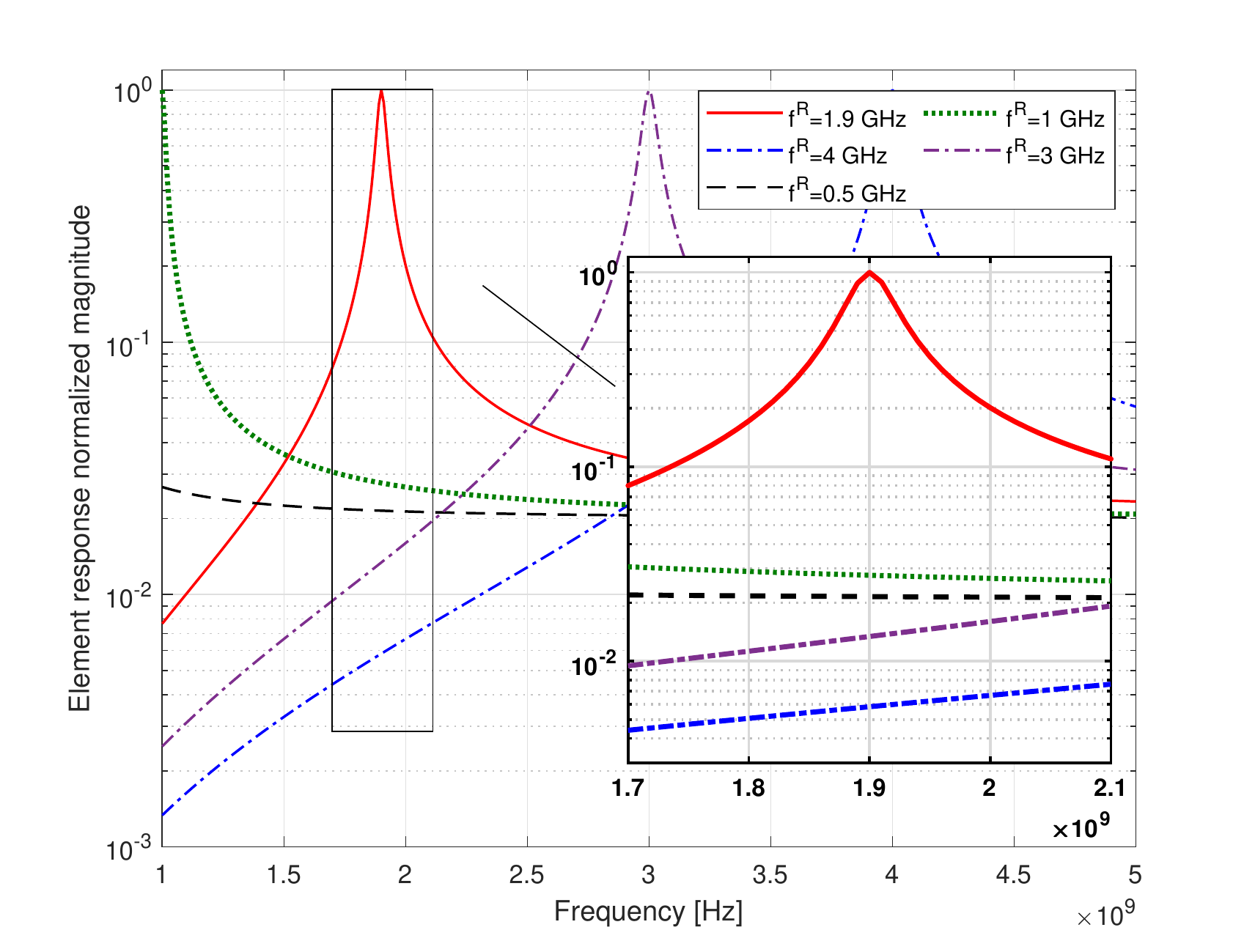}}
			\vspace{-0.8cm}
			\caption{Element response magnitude vs. frequency.}
			\label{fig:Lorentz1Mag}		
		\end{minipage}
		$\quad$
		\begin{minipage}{0.45\textwidth}
			\centering
			\scalebox{0.48}{\includegraphics{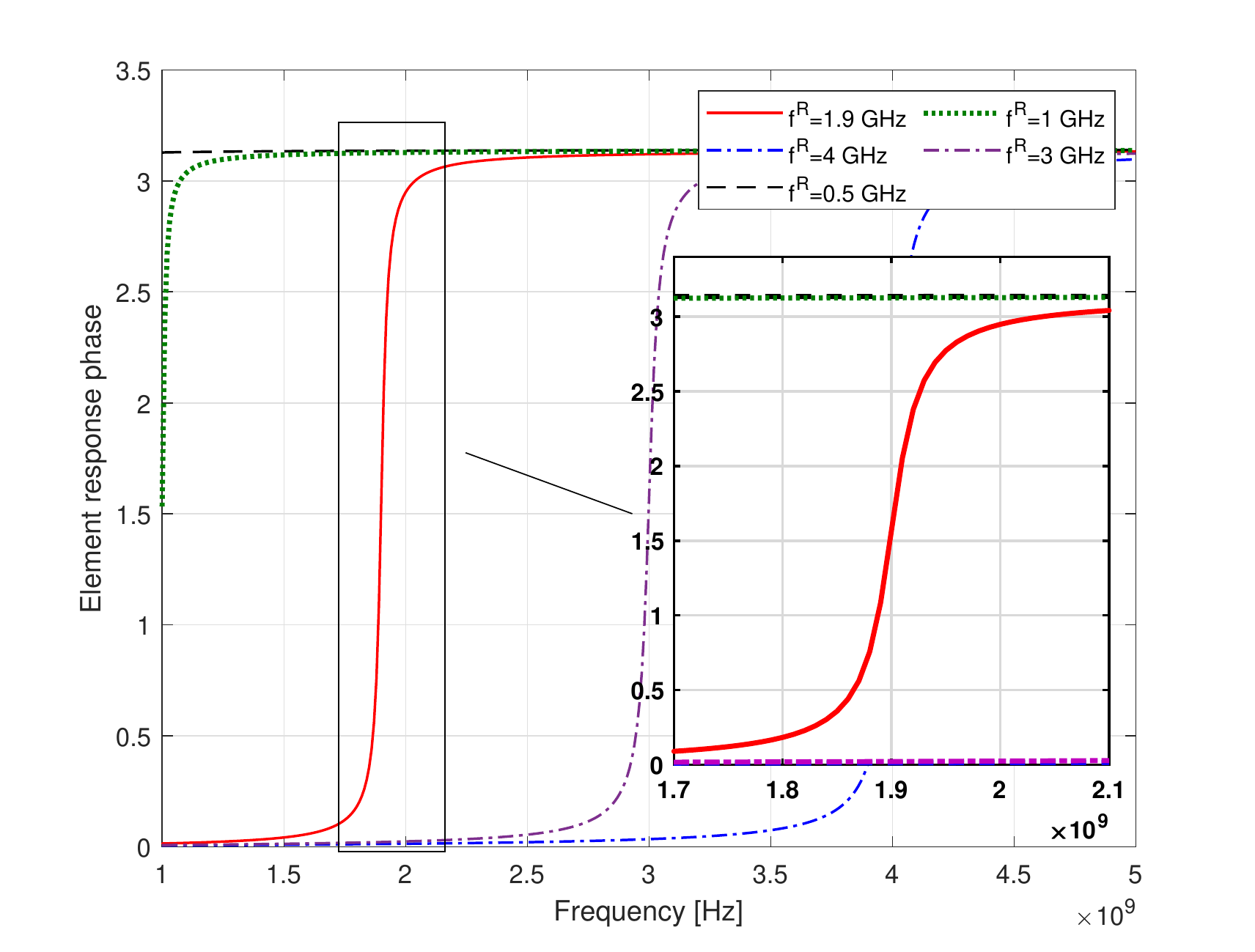}}
			\vspace{-0.8cm}
			\caption{Element response phase vs. frequency.}
			\label{fig:Lorentz1Phase}
		\end{minipage}
		\vspace{-0.2cm}
	\end{figure}

	As an example, we depict in Figs. \ref{fig:Lorentz1Mag}-\ref{fig:Lorentz1Phase} the frequency response of a single element with magnitude normalized to unity at resonance, i.e., $\frac{\alpha(f)}{\alpha(f^R)}$, in which the ratio $\frac{f^R}{\chi}$, referred to as the quality factor, is fixed to $50$. The frequency response is evaluated for several different resonance frequencies $f^R$, focusing on the frequency band around $1.9$ GHz. Observing Figs. \ref{fig:Lorentz1Mag}-\ref{fig:Lorentz1Phase}, we note that each element approximates a bandpass filter, by setting its resonance frequency to be within the observed bandwidth, or alternatively, a filter with a frequency monotonic and even a frequency flat profile, when $f^R$ is outside of the band of interest. In the sequel
	we show that this frequency selectivity can be exploited to facilitate \ac{ofdm} signal recovery in the presence of quantized measurements.
	
	%

\begin{figure}
		\centering	
		\includegraphics[scale=0.3]{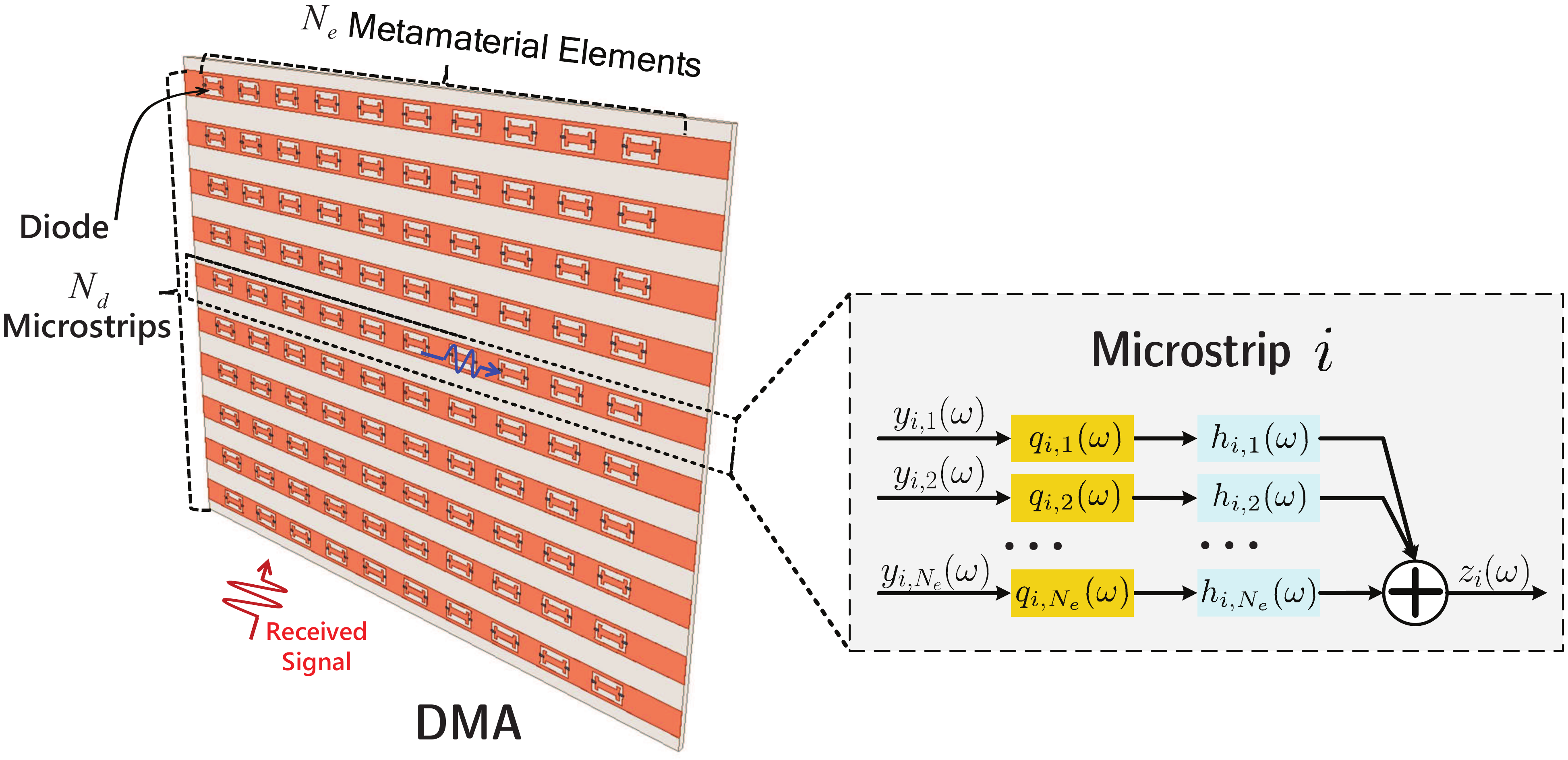}
		\vspace{-0.4cm}
		\caption{DMA model illustration.} 
		\label{fig:MetaAntenna}
	\end{figure}

	Processing of the \ac{dma} output is carried out in digital baseband. Therefore, we next formulate the resulting model in discrete-time. Consider a \ac{dma} with $N\triangleq N_d \cdot N_e$ tunable metamaterial elements, where $N_d$ and $N_e$ are the number of microstrips and elements in each microstrip, respectively.
	Let $y_{i,l}[t]$ denote the equivalent baseband signal received from the wireless channel on metamaterial element $l$ of microstrip $i$ at time slot $t \in \{0,1,\ldots, M-1\} \triangleq \mySet{M}$, where $M$ is the transmission block size, and let  $y_{i,l}(\omega)$ be its \ac{dtft}.
	The frequency domain representation of the output of the $i$th microstrip, dentoed $z_i(\omega)$ can be written as
	\begin{equation}\label{eqn:DMO_Rel1}
	z_i(\omega) = \sum\limits_{l=1}^{N_e}h_{i,l}(\omega)q_{i,l}(\omega)y_{i,l}(\omega),
	\end{equation} 
	where $h_{i,l}(\omega)$ characterizes the effect of the signal propagation inside the microstrip \ref{itm:P2}, while $q_{i,l}(\omega)$ denotes the tunable weight of the $l$th element of the $i$th microstrip. Based on property \ref{itm:P1}, this frequency dependent profile, which represents the \ac{dtft} equivalent of \eqref{eqn:FreqSel}, is given by
	\begin{equation}\label{eqn:Form_of_Weight}
	q_{i,l}(\omega)=\frac{F_{i,l} \Omega^{2}(\omega)}{(\Omega_{i,l}^{R})^{2}-\Omega^{2}(\omega)-j \Omega(\omega) \chi_{i,l}}.
	\end{equation}	
	In the above equation, $F_{i,l}$, $\chi_{i,l}$, and $\Omega_{i,l}^{R}$ are the configurable  oscillator strength, damping factor, and angular resonance frequency, respectively, of the $l$th element of microstrip $i$. We use $\Omega(\omega)$ to denote the analogous angular frequency corresponding to the continuous-time frequency $\omega$, namely, how the bandwidth of interest in continuous-time is mapped into angular frequencies of the \ac{dtft} of the discrete-time signal. This mapping  is dictated by the carrier frequency $f_c$ and the sampling rate $f_s$, and can be written as
	\[\Omega (\omega ) =
	{2\pi {f_c} + \omega {f_s}}, \qquad { |\omega|  < \pi }
	.\]
	An illustration of the system is given in  Fig. \ref{fig:MetaAntenna}.
	
	Next, we express the \ac{dma} model in \eqref{eqn:DMO_Rel1} compactly in vector form.
	To that aim, we define a set of $N\times N$ diagonal matrices $\qH(\omega)$ whose $[\left( {i - 1} \right){N_e} + l]$th diagonal element is $h_{i,l}(\omega)$. 
	We also define  $\qy(\omega)\in\calC^{N}$ as the vector comprised of the frequency domain received signal of the complete array such that its $[\left( {i - 1} \right){N_e} + l]$th element is $y_{i,l}(\omega)$. 
	Using these definitions and letting $\qz(\omega)\in\calC^{N_d}$ be the \ac{dtft} of the \ac{dma} output, i.e., $(\qz(\omega))_i = z_i(\omega)$, the frequency domain formulation of the \ac{dma} operation can be formulated as
	\begin{equation}\label{org_channel}
	{\qz}(\omega)=\qQ(\omega)\qH(\omega)\qy(\omega),
	\end{equation} 
	where the matrix $\qQ(\omega)\in\calC^{N_d\times N}$ represents the tunable weights, and its entries are given by $\left(\qQ(\omega)\right)_{k, \left({i - 1} \right){N_e} + l}={{q_{i,l}(\omega)}}$ when $i = k$, and $0$ when $i\ne k$, where $l 
	\in \{1,2, \cdots {N_e}\}\triangleq \mathcal{N}_e$ and $i,k \in \{1,2, \cdots {N_d}\}\triangleq \mathcal{N}_d$.
	By its definition, $\qQ(\omega)$ is block diagonal with diagonal blocks being row vectors denoted by $\big({\qq_i(\omega)}\big)^{T}=[q_{i,1}(\omega),q_{i,2}(\omega),\cdots,q_{i,N_e}(\omega)]\in\calC^{N_e}$ for $i\in\mathcal{N}_d$.
	The vector expression \eqref{org_channel} is utilized  to formulate bit-constrained \ac{dma}-based \ac{mu}-\ac{mimo}-\ac{ofdm} systems, as detailed in the next subsection.
	
	\subsection{Received Signal Model}
	\label{subsec:MIMO_OFDM}
	
	In this paper, we consider the uplink scenario of a single-cell \ac{mu}-\ac{mimo}-\ac{ofdm} system. 
	Here, the \ac{bs} is equipped with a \ac{dma}  consisting of $N_d$ microstrips with a total of $N=N_e \cdot N_d$ metamaterial elements, and serves $K$ single-antenna users. 
	The users simultaneously transmit \ac{ofdm} symbols with  $M$   subcarriers each. 
	Due to power or memory constraints, the \ac{bs} processes a coarsely quantized version of the \ac{dma} output obtained using \acp{adc}, modeled as identical uniform scalar quantizers with resolution $b$. 
	In particular, we focus on the recovery of the transmitted \ac{ofdm} symbols in a hybrid manner by jointly configuring the \ac{dma} weights along with a digital filter applied to the \acp{adc} outputs.
	
	We consider a frequency-selective wireless channel, which follows a tapped delay line model with $L_{G}$ taps, represented by a set of $K\times N$ matrices $\{\qG[\tau]\}_{\tau=0}^{L_{G}-1}$.
	Let $\qs[t] \in \mathcal{C}^{K}$ be the \ac{ofdm} symbols transmitted at the $t$th time slot, assumed to be i.i.d. and with covariance matrix $\qI_{K}$. We focus on time instances $t \in \mySet{M}$, which correspond to the \ac{ofdm} block after \ac{cp} removal. 
	We denote the \ac{dtft} of $\{\qG[\tau]\}_{\tau=0}^{L_{G}-1}$ and $\{\qs[t]\}_{t=0}^{M-1}$ as $\qG(\omega)$ and $\qs(\omega)$, respectively.
	Then, the \ac{dtft} representation of 
	the received channel output vector $\qy(\omega)$ is expressed as
	\begin{equation}
	\label{eqi_channel}
	\qy(\omega)=\qG(\omega)\qs(\omega)+\qw(\omega),
	\end{equation} 
	where $\qw(\omega)\in\calC^{N}$ is the \ac{dtft} of the additive noise vectors  $\{\qw[t]\}_{t=0}^{M-1}$,  which are independent of $\qs[t]$ and follow a zero-mean proper-complex Gaussian distribution with covariance $\qC_W$ for each $\omega$. 
	Since the elements in every microstrip are commonly sub-wavelength spaced, they are typically spatially correlated, and thus $\qC_W$ is not restricted to be diagonal. 
	The combined effect of the wireless channel and the propagation inside the microstrips in the frequencty domain can be represented by the equivalent channel $\hat{\qG}(\omega)=\qH(\omega)\qG(\omega)$.
	
	
	We assume the \ac{cp} length is larger than the memory length of   the equivalent channel $\hat{\qG}(\omega)$.
	Consequently, the frequency response of all considered signals and channels is fully captured by its $M$-point \ac{dft}, i.e., the \ac{dtft} in angular frequencies $\{\omega_m \triangleq \frac{2\pi m}{M}\}_{m \in \mySet{M}}$.
	%
	For brevity, we henceforth write $\hat{\qG}_m\triangleq\hat{\qG}_m(\omega_m)$, and similarly define $\qQ_m$, $\qH_m$, $\qz_m$, $\qs_m$ and $\qw_m$.
	Using the above  notations and substituting  \eqref{org_channel} into \eqref{eqi_channel} for all $M$ subchannels, the input-output relationship of a single \ac{ofdm} block after \ac{cp} removal in the frequency domain can be written  in matrix form in which each row represents a single frequency bin via 
	\begin{equation}
	\label{eqn:DMA outputFD}
	\bar{\qZ}=\bar{\qQ}\bar{\qG}\bar{\qS}+\bar{\qQ}\bar{\qH}\bar{\qW},
	\end{equation} 
	where $\bar{\qZ}\triangleq[\qz_0^T,\qz_1^T,\cdots,\qz_{M-1}^T]^{T}$, $\bar{\qS} \triangleq[\qs_0^T,\qs_1^T,\cdots,\qs_{M-1}^T]^{T}$, and $\bar{\qW} \triangleq[\qw_0^T,\qw_1^T,\cdots,\qw_{M-1}^T]^{T}$ denote the vertically concatenated vectors of the \ac{dma} output, transmitted signal and additive noise observed on each subchannel, respectively;  
	$\bar{\qQ}\triangleq{\mathop{\rm blkdiag}\nolimits} \left( {\qQ_0, \qQ_1,\cdots ,\qQ_{M-1}}\right)$, $\bar{\qG}\triangleq{\mathop{\rm blkdiag}\nolimits} \big( {\hat{\qG}_0, \hat{\qG}_1,\cdots ,\hat{\qG}_{M-1}} \big)$ and $\bar{\qH}\triangleq{\mathop{\rm blkdiag}\nolimits} \left( {\qH_0, \qH_1,\cdots,\qH_{M-1}}\right)$ denote the block diagonal formulation of DMA weights, equivalent channel and \ac{dma} propagation characterization on each subchannel, respectively. The expression \eqref{eqn:DMA outputFD} models the \ac{dma} output, which is fed to the \acp{adc}, in the frequency domain. Since the \ac{adc} operation is formulated in the time domain, we transform $\bar{\qZ}$ into the time domain by multiplying it with $\qV_1\triangleq\left( {{\bf{F}}_{M}^H \otimes {{\bf{I}}_{N_d}}} \right)$, where $\qF_{M}$ is the $M\times M$ normalized \ac{dft} matrix. The resulting \ac{dma} output is given by:
	\begin{equation}
	\qZ=\left( {{\bf{F}}_{M}^H \otimes {{\bf{I}}_{N_d}}} \right)\bar{\qZ}=\qV_1\bar{\qQ}\bar{\qG}\qV_2{\qS}+\qV_1\bar{\qQ}\bar{\qH}\bar{\qW},
	\label{eqn:DMA outputTD}
	\end{equation} 
	where in \eqref{eqn:DMA outputTD} we used the definitions  $\qV_2\triangleq\left( {{\bf{F}}_{M} \otimes {{\bf{I}}_K}} \right)$, ${\qS}=[\qs[0]^T,\qs[1]^T,\cdots,\qs[M-1]^T]^T$,  and $\qZ=[\qz[0]^T,\qz[1]^T,\cdots,\qz[M-1]^T]^T$.
	
	%
	
	\begin{figure}
		\centering	
		\includegraphics[scale=0.5]{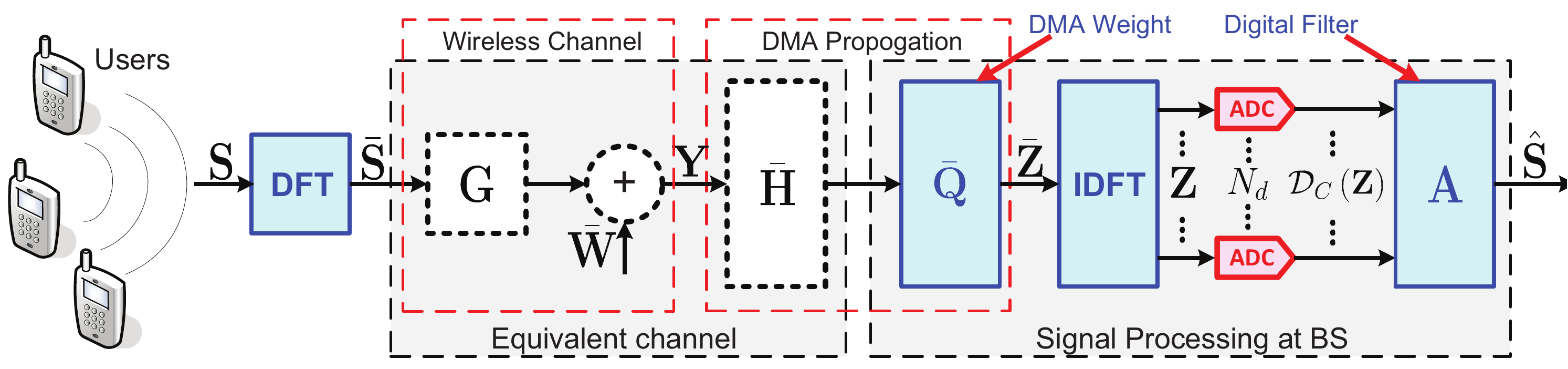}
		\vspace{-0.1cm}
		\caption{Illustration of signal processing procedure. The configurable receiver parameters are the digital filter $\qA$, the \ac{dma} weights $\bar{\qQ}$, and the \ac{adc} support $\gamma$. } 
		\label{fig:System model}
	\end{figure}
	
	The \ac{dma} outputs $\{\qz[t]\}_{t=0}^{M-1}$ are quantized using $N_d$ identical pairs of \acp{adc} which independently discretize the real and imaginary parts of each analog input sample.
	We denote the complex-valued quantization mapping by  $\calD_{C}(x+jy)=\calD(x)+j\calD(y)$, where $\calD(\cdot)$ is the uniform real-valued quantization operator with support $\gamma$, i.e.,  
	\vspace{-0.1cm}
	\begin{equation}
	\calD(x) =   \begin{cases}
	-\gamma + \frac{2\gamma}{b}\left(l + \frac{1}{2} \right)   & x - l \cdot \frac{2\gamma}{b} + \gamma \in  \left[0,\frac{2\gamma}{b} \right], l \in \{0,1, \ldots, b - 1 \},\\
	{\rm sign}\left( x\right) \left( \gamma - \frac{\gamma}{b}\right)    & |x| > \gamma.
	\end{cases}
	\vspace{-0.1cm}
	\label{eqn:Qrule}
	\end{equation}  
	The recovered symbols are obtained by linearly processing the quantized \ac{dma} outputs.
	Letting the $\qA$ represent the digital linear processing, the resulting estimate of $\qS$ is  expressed by 
	\vspace{-0.1cm}
	\begin{equation}
	\label{eqn:Shat}
	{\bf{\hat S}} = \qA\calD_{C}\left( \qZ \right).
	\vspace{-0.1cm}
	\end{equation}
	An illustration of the overall system is depicted in Fig. \ref{fig:System model}. It is noted that resulting formulation models the processing of the received signal as a bit-constrained hybrid system with frequency selective analog combining, represented by the weights $\bar{\qQ}$. As such, the system in Fig. \ref{fig:System model} specializes a broad range of conventional hybrid receiver architectures, in which the analog processing is frequency flat, e.g., \cite{mendez2016hybrid,ioushua2019family,zhang2005variable}. For example, this model specializes bit-constrained hybrid receivers with partially-connected  phase shifter networks, as considered in \cite{roth2018comparison}, by fixing $\qH(\omega)$ to be the identity matrix, i.e., canceling the effect of the propagation inside the microstrips \ref{itm:P2}, while restricting the elements $q_{i,l}(\omega)$ of which $\bar{\qQ}$ is comprised to be independent of $\omega$ and have a unit magnitude, namely, $q_{i,l}(\omega) \equiv e^{j\phi_{i,l}}$ for some $\phi_{i,l}\in [0,2\pi)$ for all $(i,l) \in \calN_d \times \calN_e$. 
	
	Our aim is to jointly design the \ac{dma} weights $\bar{\qQ}$,  the digital linear filter $\qA$, and the \ac{adc} support $\gamma$ to minimize the symbol recovery \ac{mse} $\E\{\|{\qS} -\hat{\qS} \|^2\}$, namely to produce an accurate estimate of ${\qS}$. 
	The  estimator $\hat{\qS}$ can be then used to facilitate the detection of each symbol, taking values in some discrete constellation.  
	
	\section{DMA Design}
	\label{sec:DMA}
	We now propose  a \ac{dma} configuration scheme to minimize the signal recovery \ac{mse}. Our method is derived by treating the joint design of the \ac{dma} weights  and the digital filter  as a task-based quantization setup \cite{shlezinger2018hardware}.  
	We first formulate the problem accordingly in Subsection~\ref{subsec:TaskBased}. Then, in Subsection~\ref{subsec:Greedy} we propose a greedy method for configuring the \ac{dma} weights assuming that they can take any frequency selective profile. Next, we show in Subsection \ref{subsec:Project} how these unconstrained weights can be used to choose a \ac{dma} with approximately frequency flat weights, as assumed in \cite{shlezinger2019dynamic}, as well as elements obeying the more general Lorentzian frequency selective profile, as stated in property \ref{itm:P1}. Finally, we discuss some of the insights and related aspects which arise from these designs in Subsection~\ref{subsec:Discussion}.

	\subsection{Task-Based Quantization Formulation}
	\label{subsec:TaskBased}
	Task-based quantization refers to the joint optimization of the quantization rule along with its pre and post quantization mappings in light of a given system task \cite{shlezinger2018hardware,shlezinger2018asymptotic,salamatian2019task,shlezinger2019deep}. 
	The bit-constrained receiver architecture detailed in Subsection \ref{subsec:MIMO_OFDM} can be treated as a task-based quantization setup, since the output of the wireless channel is acquired and discretized for the task of recovering the transmitted \ac{ofdm} symbols. In particular, the formulations of the \ac{dma} output in \eqref{eqn:DMA outputTD} and the estimated symbols in \eqref{eqn:Shat} indicate that the \ac{dma} induces a configurable pre-quantization mapping, represented by the matrix $\qV_1\bar{\qQ}$, while the  filter $\qA$ corresponds to the digital post-quantization processing. Using this framework, we next characterize the achievable \ac{ofdm} recovery accuracy for a given \ac{dma} configuration, which is used in the following subsection to derive algorithms for setting the \ac{dma} weights.  
	
	Following \cite{shlezinger2018hardware}, we derive $\bar{\qQ}$ and $\qA$  assuming that the \acp{adc} are not overloaded, i.e., that the magnitudes of the real and imaginary parts of $\qZ$ are not larger than the \ac{adc} support $\gamma$, with sufficiently large probability.
	To guarantee this, we set $\gamma$ to be some multiple $\eta>0$ of the maximal standard deviation of the \ac{adc} inputs, such that the overload probability is bounded via Chebyshev's inequality \cite[Pg. 64]{cover2012elements}. 
	This setting can be expressed as
	\begin{align}
	\gamma^2 &= {\eta^2} \max_{t\in\calM}\max_{i\in\calN_d}{\E\{|(\qz[t])_i|^2 \}} \notag \\
	&\stackrel{(a)}{=} 	{{\eta^2} \max_{i\in\calN_d}\frac{1}{M}\sum_{t\in\calM}{\E\{|(\qz[t])_i|^2 \}} \stackrel{(b)}{=} {\eta^2} \max_{i\in\calN_d}\frac{1}{M}\sum_{m\in\calM}{\E\{|(\qz_m)_i|^2 \}}} \notag \\
	&= {\eta^2}\max_{i\in\calN_d}\frac{1}{M}\sum_{m\in\calM}{\qq}_{m,i}^T\qE^T_i\qUpsilon_m\qE_i{\qq}_{m,i}^{*},
	\label{eqn:Support}
	\end{align}
	where $(a)$ follows from the stationarity of baseband \ac{ofdm} symbols after \ac{cp} removal \cite{heath1999exploiting}, and $(b)$ is obtained from Parseval's equality. In \eqref{eqn:Support},   $\qE_i$ represents the selection of the microstrip, and is given by $\qE_i=[\qE_{i,1},\qE_{i,2},\cdots,\qE_{i,N_d}]^T$ with $\qE_{i,i}\!=\! \qI_{N_e}$   while $\qE_{i,j}\!=\!\qO_{N_e}$ for $j\neq i$. The matrix $\qUpsilon_m$ is the covariance of the equivalent channel output at frequency bin $m\in\calM$, given by $\qUpsilon_m\triangleq\hat{\qG}_m\hat{\qG}^H_m+\qH_m\qC_W\qH^H_m$.
	The \ac{mse}-minimizing digital filter for a fixed $\bar{\qQ}$ with non-overloaded \acp{adc}, and its corresponding \ac{mse}  for asymptotically large $M$ are stated in the following lemma:
	\begin{lemma}
		\label{lem:taskBased}
		The \ac{mse} minimizing digital filter $\qA$ under a fixed \ac{dma} configuration $\bar{\qQ}$ and \ac{adc} support $\gamma$ is given by 
		\begin{equation}\label{Mat_min_mmse}
		\qA^{*}(\bar{\qQ})\!=\!\qV_2^{H}\bar{\qG}^H\bar{\qQ}^H{{\left(  {\sigma^2_q}{{\bf{I}}_{M{N_d}}} \!+\! {\bar{\qQ}\qSigma\bar{\qQ}^H} \right)}^{ - 1}}\qV_1^{H}\!,
		\end{equation}
		where  $\sigma^2_q\triangleq\frac{4\gamma^2}{3b^2}$ represents the quantization noise energy, and 
		{$\qSigma\triangleq {\rm blkdiag}\big(\qUpsilon_0,\qUpsilon_1,\dots, \qUpsilon_{M-1} \big)$}. 
		The corresponding  \ac{mse} is  	 $\mathrm{EMSE}(\bar{\qQ})+e^{\rm o}$, where 
		$e^{\rm o}$ is the \ac{mmse} in recovering ${\qS}$ from the channel output, and the excess \ac{mse} is
		\begin{equation}\label{min_mmse} 
		{\mathrm{EMSE}(\bar{\qQ})\!=\!\tr\left[ {\bar{\qG}^H\qSigma^{-1}{{\left( {\qSigma^{-1} \!+\! \sigma^{-2}_q\bar{\qQ}^H\bar{\qQ}} \right)}^{ \!-\! 1}}\qSigma^{-1}\bar{\qG}} \right].}
		\end{equation}
		
	\end{lemma}
	
	\ifproofs	
\begin{IEEEproof}
    The asymptotic Gaussianity of \ac{ofdm} signals \cite{wei2010convergence} implies that the \ac{mmse} estimate  of ${\qS}$ is linear when the number of subcarriers  $M$ is asymptotically large.	Consequently, in that regime, the linear estimator  of ${\qS}$  from the channel output achieves the \ac{mmse} $e^{\rm o}$, and thus the problem of identifying $\qA^{*}(\bar{\qQ})$ and its corresponding \ac{mse} is a special case of \cite[Lemma 1]{shlezinger2018hardware}, formulated for complex-valued signals. 
\end{IEEEproof}
	\fi	 	
	
	It follows from the proof of Lemma \ref{lem:taskBased} that the requirement of $M$ to be asymptotically large is necessary to guarantee that the \ac{mmse} estimate of  ${\qS}$ from the channel output is linear, namely, that the transmitted \ac{ofdm} symbols approach being Gaussian. Consequently, the lemma also rigorously holds for finite $M$ when the entries of $\qS$ obey a jointly Gaussian distribution, i.e., the transmitters utilize Gaussian symbols, as commonly assumed in the massive \ac{mimo} literature \cite{Marzetta-2010TWC,hoydis2013massive, shlezinger2018spectral}. 
	
	Lemma \ref{lem:taskBased} characterizes the achievable \ac{mse} of the \ac{dma}-based \ac{bs} in recovering the \ac{ofdm} signals matrix $\qS$ for a given \ac{dma} weights $\bar{\qQ}$. 
	While the excess \ac{mse} in \eqref{min_mmse} holds rigorously for non-overloaded \acp{adc} and asymptotically large number of subcarriers, it also constitutes a close approximation of the error  for systems with a finite number of subcarriers and a small but not necessarily zero probability of overloading the quantizers under a broad range of channel output distributions \cite{shlezinger2018hardware}, and is thus used in the following as an objective function for setting the \ac{dma} weights.
	Nonetheless, determining the feasible \ac{dma} configuration which minimizes the excess \ac{mse} in  \eqref{min_mmse} is a challenging task.
	Particularly, the constraints on the structure of  $\bar{\qQ}$ and the feasible values of its entries, as well as the expression of the objective  function \eqref{min_mmse}, make the derivation of a tractable closed-form expression for the \ac{mse}-minimizing \ac{dma} weights difficult.
	However, the problem can be simplified by accounting for the block diagonal structure of $\{\qQ_m\}_{m=0}^{M-1}$ and $\qSigma$.
	Specifically, 
	\ifproofs
	$\qQ_m^H$ is a block diagonal matrix of column vectors $\{{\qq}_{m,i}^{*}\}_{i=1}^{N_d}$, namely, $\qQ_m^H={\mathop{\rm blkdiag}\nolimits} \left( {{\qq}_{m,1}^{*}, {\qq}_{m,2}^{*},\cdots ,{\qq}_{m,N_d}^{*}}\right)$.
	As a result, 
	\fi
	$\qQ_m^H\qQ_m$ is also a block diagonal matrix comprised of the rank-one submatrices $\{{\qq}_{m,i}^{*}{\qq}_{m,i}^{T}\}_{i=1}^{N_d}$ along its main diagonal.
	This structure leads to the following formulation of the excess MSE: 
	\ifproofs	
	\else
	Proposition \ref{pro:BlkDiagStructure}, whose proof is omitted due to space limitations:
	\fi	
	\begin{proposition}
		\label{pro:BlkDiagStructure}
		The excess \ac{mse} \eqref{min_mmse} can be written as
		\vspace{-0.1cm}
		\ifsingle
		\begin{equation}\label{min_mmse_simp}
		\mathrm{EMSE}(\bar{\qQ})=\sum_{m=1}^{M}\tr\bigg[\hat{\qG}_{m}^H\qUpsilon_m^{-1}\Big(\qUpsilon_m^{-1} +\sigma^{-2}_q\sum_{i=1}^{N_d}\qE_i\qq_{m,i}^{*}{\qq}^{T}_{m,i}\qE^T_i\Big)^{ - 1}\qUpsilon_m^{-H}\hat{\qG}_{m}\bigg].
		\vspace{-0.1cm}
		\end{equation}
		\else
		\begin{equation}\label{min_mmse_simp}
		\!	\!\mathrm{EMSE}(\bar{\qQ})\!=\!\sum_{m=1}^{M}\!\tr\bigg[\hat{\qG}_{m}^H\qK_{m}\!\Big(\!\qK_{m} \!+\!\sigma^{-2}_q\sum_{i=1}^{N_d}\qE_i\qq_{m,i}^{*}{\qq}^{T}_{m,i}\qE^T_i\Big)^{\! - 1}\!\!\!\!\qK_{m}^H\hat{\qG}_{m}\bigg].
		\vspace{-0.1cm}
		\end{equation}
		\fi		
	\end{proposition}
	
	\ifproofs	
	{\em Proof:}
	See Appendix \ref{app:Proof1}.
	\fi	 
	
	The alternative formulation of $\mathrm{EMSE}(\bar{\qQ})$ allows to optimize the vectors $\{\qq_{m,i} \}$ individually in a greedy manner, instead of the matrix $\bar{\qQ}$ directly.
	Based on this strategy, we next propose a \ac{dma} configuration algorithm.

	\vspace{-0.2cm}
	\subsection{Greedy Unconstrained \ac{dma} Configuration}
	\label{subsec:Greedy}
	\vspace{-0.1cm} 
	We next propose a DMA weight design strategy  based on the objective \eqref{min_mmse_simp}. We note that recovering the weights which minimize  \eqref{min_mmse_simp} is difficult due to the constrained feasible set, modeled via property \ref{itm:P1}, and the fact that   the excess \ac{mse} \eqref{min_mmse_simp}  is not separable in the microstrip index $i \in \calN_d$. 
	Therefore, our design approach consists of two steps: First, we propose a greedy method for optimizing the weights individually for each sub-channel $m\in\mathcal{M}$, assuming unconstrained weights, i.e., that the frequency response can be set for each element at each frequency bin individually. Then, we show how these unconstrained weights can be approximated using a feasible \ac{dma} configuration in the following subsection.

	When the \ac{dma} weights can be tuned individually in each frequency bin, the problem of configuring the elements is formulated as
	\ifsingle
	\begin{align} 
	\{\qq_{m,i}\}_{i=1}^{N_d}  = \mathop{\arg\min}\limits_{\{{\bar{\qq}}_{i}\in\calC^{N_e}\}} \tr\bigg[\hat{\qG}_{m}^H\qUpsilon_m^{-1}\Big(\qUpsilon_m^{-1} +\sigma^{-2}_q\sum_{i'=1}^{N_d}\qE_{i'}\bar{\qq}_{i'}^{*}{\bar{\qq}}^{T}_{i'}\qE^T_{i'}\Big)^{ - 1}\qUpsilon_m^{-H}\hat{\qG}_{m}\bigg]
	\label{min_mmse_sub}. 
	\end{align}	
	\else
	\begin{align}
	\small
	\!\!\{\qq_{m,i}\} \!=\!\mathop{\arg\min}\limits_{\{{\bar{\qq}}_{i}\in\calC^{N_e}\}}\!\tr\bigg[\!\qP_{m}\! &\Big(\qK_{m} \!\!+\!\sigma^{\!-\!2}_q\!\sum_{j\!=\!1}^{N_d}\qE_j\bar{\qq}_j^{*}\bar{\qq}^{T}_j\qE^T_j\Big)^{ \!-\! 1} \!\! \qP^H_{m}\bigg]
	\label{min_mmse_sub}.
	\vspace{-0.1cm}
	\end{align}
	\fi
	Due to the difficulty in optimizing \eqref{min_mmse_sub} jointly for the weights of all the microstrips, i.e., over all $i\in\calN_d$, we  set the weights associated with each micropstrip separately, updating the complete DMA in a sequential manner.  
	In particular, the method operates iteratively, where in the $i$th iteration, we account for the contribution of the $i$th microstrip to the \ac{mse}  and optimize its  weight vector ${\qq}_{m,i}$ given the  previously designed weights $\{{\qq}_{m,j}\}_{j=1}^{i-1}$. 
	To formulate the greedy method,  define $\qU_{m,0}\triangleq\qUpsilon_m^{-1}$ and $\qU_{m,i} \triangleq \qUpsilon_m^{-1} + \sigma^{-2}_q\sum_{j = 1}^{i} {{{\bf{E}}_j}\qq_{m,j}^{*}{\qq}^{T}_{m,j}{\bf{E}}_j^T}$ for $i\geq1$, which can be written as $\qU_{m,i}=\qU_{m,i-1}+\sigma^{-2}_q\qE_i\qq_{m,i}^{*}{\qq}^{T}_{m,i}\qE_i^H$.
	The MSE in \eqref{min_mmse_sub} can now be computed by recursively evaluating 
	\vspace{-0.1cm}
	\begin{align}
	J_{m,i}&=\tr\left[ \hat{\qG}_{m}^H\qUpsilon_m^{-1}{{\left( \qU_{m,i-1}+\sigma^{-2}_q\qE_i\qq_{m,i}^{*}{\qq}^{T}_{m,i}\qE_i^T \right)}^{ - 1}}\qUpsilon_m^{-1}\hat{\qG}_{m} \right]\notag\\
	&\stackrel{(a)}{=}\tr\left[\hat{\qG}_{m}^H\qUpsilon_m^{-1}\qU_{m,i-1}^{-1}\qUpsilon_m^{-1}\hat{\qG}_{m}\right]-\frac{{\qq_{m,i}^T\qXi_{m,i}{\qq_{m,i}^{*}}}}{\sigma^{2}_q+\qq_{m,i}^T\qPsi_{m,i}\qq_{m,i}^{*}},\label{J_i}
	\vspace{-0.1cm} 
	\end{align}
	where $(a)$ follows from the Sherman-Morisson formula \cite[Ch. 3.8]{meyer2000matrix},
	and we define $\qXi_{m,i}\triangleq{\bf{E}}_i^T{\bf{U}}_{m,i - 1}^{ - 1}{\qUpsilon_m^{-1}}\hat{\qG}_{m}\hat{\qG}_{m}^H{\qUpsilon_m^{-1}}{\bf{U}}_{m,i - 1}^{ - 1}{{\bf{E}}_i}$ and $\qPsi_{m,i}\triangleq{\bf{E}}_i^T{\bf{U}}_{m,i - 1}^{ - 1}{{\bf{E}}_i}$. 
	Note that  $J_{m,i}$ in \eqref{J_i} coincides with the objective in \eqref{min_mmse_sub} for $i=N_d$.
	Our proposed greedy approach sequentially selects ${\qq}_{m,i}$ that minimizes $J_{m,i}$, which is equivalent to maximizing the second term of \eqref{J_i}, given that $\{{\qq}_{m,j}\}_{j=1}^{i-1}$ have been determined in the previous iterations, i.e.,  that $\qU_{m,i-1}$ is known.
	Note that the quantization noise energy $\sigma_q^2$, which depends on the ADC support $\gamma$, is determined by the complete matrix $\bar{\qQ}$ via \eqref{eqn:Support}.
	To facilitate the minimization of $J_{m,i}$ with respect to ${\qq}_{m,i}$ for every $i\in \calN_d$, we optimize each ${\qq}_{m,i}$ assuming that $\gamma$ in $\sigma^{2}_q$ is dictated by the output of the $i$th microstrip at the $m$th frequency bin, i.e., $\gamma^2=\eta^2 \E\{|(\qz_{m})_i|^2 \}$.
	By defining $\kappa \triangleq \frac{4\eta^2}{3b^2}$ and substituting \eqref{eqn:Support} into \eqref{J_i}, the resulting optimization problem becomes
	\vspace{-0.1cm}
	\begin{equation}
	\label{eqn:ProbForm2} 
	{\qq}_{m,i} =
	\mathop{\arg\max}\limits_{{\qq}\in\calC^{N_e}}\quad\bar{J}_{m,i}({\qq})\triangleq\frac{{{\bf{ q}}^T\qXi_{m,i}{{{\bf{ q}}}^{*}}}}{{{\bf{ q}}^T\left(\kappa\qE^T_i\qUpsilon_m\qE_i+\qPsi_{m,i}\right){{{\bf{ q}}}^{*}}}}.
	\vspace{-0.1cm} 
	\end{equation} 
	%
	The solution of \eqref{eqn:ProbForm2} is characterized in the following lemma:

	\begin{lemma}
		\label{lem:unconstrained}
		The solution to \eqref{eqn:ProbForm2} 
		is given by $\alpha_{m,i} \hat{\qq}_{m,i}$ for any $\alpha_{m,i} \in \mySet{C}$, where  $\hat{\qq}_{m,i}^*$ is the   generalized eigenvector corresponding to the maximal generalized eigenvalue  of $\qXi_{m,i}$ and  $\kappa\qE^T_i\qUpsilon_m\qE_i+\qPsi_{m,i}$.		
	\end{lemma}
	
	\begin{IEEEproof}
		This lemma follows from \cite[Sec 4.5]{Ghojogh-2019Arxiv}.	
	\end{IEEEproof}	
	
	Lemma \ref{lem:unconstrained} indicates that when $\kappa\qE^T_i\qUpsilon_m\qE_i+\qPsi_{m,i}$ is invertible, then  $\hat{\qq}_{m,i}$ is the conjugate of the eigenvector corresponding to the largest eigenvalue of  $\left(\kappa\qE^T_i\qUpsilon_m\qE_i+\qPsi_{m,i}\right)^{-1}\qXi_{m,i}$. The resulting greedy algorithm is summarized as Algorithm \ref{A0}. 
	The fact that the solution in Lemma~\ref{lem:unconstrained} is invariant to the value of $\alpha_{m,i}$ is exploited to facilitate its projection into a feasible DMA weight vector, as detailed next. 
	
	\begin{algorithm} 
		\SetKwInOut{Input}{Input}
		\caption{Greedy unconstrained DMA configuration\label{A0}}
\ifsingle
\else
		\small
\fi
		\Input{Channel parameters $\qUpsilon_m$ and $\hat{\qG}_m$ for $m\in\calM$; \newline 
			\ac{adc} parameter $\kappa$.}
		\KwData{ $\qU_{m,0} = \qUpsilon_m^{-1}$ for each  $m\in\calM$.}
		\For{$i=1,2,\ldots,N_d$}{
			\For{$m=1,2,\ldots,M$}{
				Set $\hat{\qq}_{m,i}$ from $\qU_{m,i-1}$ using Lemma \ref{lem:unconstrained}\;
				Set $\qU_{m,i}\!=\!\qU_{m,i-1}\!+\!(\kappa\hat{\qq}^{T}_{m,i}\qE^T_i\qUpsilon_m\qE_i\hat{\qq}_{m,i}^{*})^{-1}\qE_i\hat{\qq}_{m,i}^{*}\hat{\qq}^{T}_{m,i}\qE_i^T$\;
			} 
		}
		\KwOut{Unconstrained weights $\{\hat{\qq}_{m,i} \}$, $(m,i)\!\in\!\calM
			\!\times\!\calN_d$.}
	\end{algorithm}
	
	\vspace{-0.2cm}
	\subsection{Setting Feasible \ac{dma} Weights}
	\label{subsec:Project}
	\vspace{-0.1cm} 
	Here, we approximate the unconstrained configuration computed via Algorithm \ref{A0}, denoted  $\{\hat{\qq}_{m,i} \}$, using a feasible \ac{dma} setting, i.e., one whose elements satisfy the model detailed in Subsection \ref{subsec:DMA_model}. In particular, we wish to set the \ac{dma} weights to minimize the distance of each element response from its corresponding unconstrained value, exploiting the invariance of each microstrip weights to a scalar factor, noted in Lemma \ref{lem:unconstrained}. Following the frequency selective element model \eqref{eqn:Form_of_Weight}, and defining $\Omega_m \triangleq \Omega(\omega_m)$, the resulting optimization problem can be expressed as the following non-linear least squares formulation:
	\begin{equation}\label{eqn:ProbProjFreq}
	\{\hat{F}_{i,l}, \hat{\chi}_{i,l},\hat{\Omega}_{i,l}^{R}\}\!=\!\mathop{\arg\min}\limits_{\{{F}_{i,l}, {\chi}_{i,l},{\Omega}_{i,l}^{R}\}} \mathop{\min}\limits_{\{\alpha_{m,i}\}}\sum_{i=1}^{N_d} \sum_{l=1}^{N_e} \sum_{m=1}^{M}\Big|\frac{F_{i,l} \Omega_m ^2}{(\Omega_{i,l}^{R})^{2}-\Omega_m ^2-j \Omega_m  \chi_{i,l}}\!-\!\alpha_{m,i}\big( \hat{\qq}_{m,i}\big)_l\Big|^2.
	\end{equation}
	Directly solving \eqref{eqn:ProbProjFreq} is a challenging task due to its non-convex structure and the fact that the objective is not separable in the frequency index $m\in\calM$.  Therefore, in the following we first seek a feasible approximation   assuming that the weights are fixed to be frequency flat, as in \cite{shlezinger2019dynamic}. Then, we show how the method can be extended to properly tune the frequency selective profile of each element as in \eqref{eqn:ProbProjFreq}. 
	
	\subsubsection{Frequency-Flat Weights} 
	Frequency flat weights is an approximation for the element frequency response which holds under narrowband signals, or alternatively, when the resonance frequency is far from the band of interest. In such cases, elements exhibit the same frequency response which takes values in some set $\mySet{Q}$, for all considered frequency bins, e.g., $\mySet{Q} = [a_{\min}, a_{\max}]$ for amplitude-only weights. Under this approximation, we denote the weights of each microstrip of index $i \in \calN_d$ by a column vector $\qq_i \in \calQ^{N_e}$, which is designed to approximate the unconstrained solutions $\{\hat{\qq}_{m,i}\}$ for each microstrip separately by solving 
	\begin{equation}\label{eqn:ProbProj}
	\qq_i=\mathop{\arg\min}\limits_{\qq\in\calQ^{N_e}}\mathop{\min}\limits_{\{\alpha_{m,i}\}_{m=1}^{M}}\sum_{m=1}^{M}\|\qq-\alpha_{m,i}\hat{\qq}_{m,i}\|^2, \qquad i \in \calN _d.
	\vspace{-0.1cm}
	%
	\end{equation}
	We tackle the optimization  problem \eqref{eqn:ProbProj}  in an alternating manner based on the following lemma: 
	\begin{lemma}
		\label{lem:Sol_Projection2}
		For a fixed $\qq_i \in \calC^{N_e}$, \eqref{eqn:ProbProj}  is minimized by setting 
		$\alpha_{m,i} = \frac{\qq_i^T \hat{\qq}^{*}_{m,i}}{\|\hat{\qq}_{m,i} \|^2}$ for all $m \in \calM$. 
		Additionally, for fixed $\{\alpha_{m,i}\}$, \eqref{eqn:ProbProj} can be solved element-wise for each $l \in \calN_e$ via 
		\begin{align}
		(\qq_i)_l &=\mathop{\arg\min}\limits_{q\in\calQ} \sum_{m=1}^{M}|q-\alpha_{m,i}(\hat{\qq}_{m,i})_l|^2.	 
		\label{eqn:Sol_Projection}
		\end{align}
	\end{lemma}
	
	\begin{IEEEproof}
	The Lemma is obtained by repeating the arguments used in \cite[Appendix B]{shlezinger2019dynamic}. 
	\end{IEEEproof}

	For $\calQ = \calC$, the value $q\in\calQ$ which minimizes \eqref{eqn:Sol_Projection} is given by $\hat{q}_{i,l} = \frac{1}{M}\sum_{m=1}^{M}\alpha_{m,i}(\hat{\qq}_{m,i})_l$. In our algorithm we thus set the entries $(\qq _i)_l$ by projecting $\hat{q}_{i,l}$ into the feasible set $\calQ$.
	Lemma~\ref{lem:Sol_Projection2} and \eqref{eqn:Sol_Projection} imply that the non-uniqueness of $\alpha_{m,i}$ can be utilized to obtain frequency-invariant feasible approximations of the frequency-selective unconstrained weights $\hat{\qq}_{m,i}$ via alternating optimization. In particular, for each microstrip index $i\in \calN_d$, the unconstrained weights are projected into a feasible set, and the resulting approximation is used to compute the weights of the remaining microstrips via the greedy method,   The  detailed procedure of the proposed DMA configuration scheme is summarized in Algorithm \ref{A1}. Once the \ac{dma} weights matrix $\bar{\qQ}$ is assigned, it is used to determine the \ac{adc} support $\gamma$ and the digital filter via \eqref{eqn:Support} and \eqref{Mat_min_mmse}, respectively.

	\begin{algorithm} 
		\SetKwInOut{Input}{Input}
		\caption{Frequency-flat setting\label{A1}}
\ifsingle
\else
		\small
\fi
		\Input{Unconstrained weights $\{\hat{\qq}_{m,i}\}$, $m\in\calM$, $i \in \calN_d$; \newline 
			alternating iterations limit ${\rm iter}_{\max}$.}
		\KwData{$\alpha_{m,i} = 1$ for all $(m,i)\!\in\!\calM
			\!\times\!\calN_d$.}
		\For{$i=1,2,\ldots,N_d$}{ 
			\For{${\rm iter}=1,2,\ldots,{\rm iter}_{\max}$}{
				Set the elements of ${\qq}_{i}$ by projecting $\frac{1}{M}\sum_{m=1}^{M}\alpha_{m,i}(\hat{\qq}_{m,i})_l$ into $\calQ$\;
				Set $\{\alpha_{m,i}\}_{m=1}^{M} $ using Lemma \ref{lem:Sol_Projection2}\; 
				
			}
		}	
		\KwOut{${\qQ}_m=\mathop{\rm blkdiag}(\qq^{T}_1,\qq^{T}_2,\cdots,\qq^{T}_{N_d})$ for all $m \in \calM$.}
	\end{algorithm}

	\subsubsection{Frequency-Selective Weights} 	
	While recovering a frequency flat \ac{dma} configuration via Algorithm \ref{A1} is relatively simple, it does not exploit the ability of the \ac{dma} elements to tune a frequency selective profile in light of the system objective. However, directly solving the non-convex optimization problem \eqref{eqn:ProbProjFreq} to recover the parameters $\{F_{i,l}, \chi_{i,l}, \Omega_{i,l}^R\}$ of each element is a difficult task. Therefore, as in the frequency flat case, we again adopt an alternating optimization approach. However, here we also account for our understanding of what types of frequency selective profiles are realized using such metamaterial elements. 
	
	In particular, we note that the parameters $\{\alpha_{m,i}\}$ which minimize the objective \eqref{eqn:ProbProjFreq} for fixed  $\{F_{i,l}, \chi_{i,l}, \Omega_{i,l}^R\}$ are obtained using Lemma \ref{lem:unconstrained}. Similarly, for fixed $\{\chi_{i,l}, \Omega_{i,l}^R\}$ and $\{\alpha_{m,i}\}$, the oscillation strength values of the elements can be tuned based on the following lemma:
	\begin{lemma}
		\label{lem:SetGain}
		For fixed  $\{\chi_{i,l}, \Omega_{i,l}^R\}$ and $\{\alpha_{m,i}\}$, the non-negative element oscillation strength values which minimize the right hand side of \eqref{eqn:ProbProjFreq} are given by
		\begin{equation}
		\label{eqn:SetGain}
		\hat{F}_{i,l} = {{\rm Re}\left( \sum\limits_{m=1}^{M}\frac{\Omega_m ^2 \alpha_{m,i}^*\big( \hat{\qq}_{m,i}\big)_l^* }{(\Omega_{i,l}^{R})^{2}-\Omega_m ^2-j \Omega_m  \chi_{i,l}} \right)^+} \cdot{\left( \sum\limits_{m=1}^{M} \frac{\Omega_m^4}{((\Omega_{i,l}^{R})^{2}-\Omega_m ^2)^2+\Omega_m^2  \chi_{i,l}^2 }\right)^{-1} }, 
		\end{equation}
		where $(x)^+ \triangleq {\rm max}(x,0)$.
	\end{lemma}
	
	\begin{IEEEproof}
		The lemma follows from \cite[Lemma 2]{shlezinger2019dynamic}.
	\end{IEEEproof}
	
	We are now left with identifying a method for setting the parameters $\{\chi_{i,l}, \Omega_{i,l}^R\}$, which essentially control how the response of each element varies in frequency. The non-convexity of \eqref{eqn:ProbProjFreq} in those parameters implies classical  non-linear least squares curve fitting methods, e.g., the Levenberg–Marquardt algorithm \cite{marquardt1963algorithm} or gradient descent, are likely to yield a local minima, unless properly initialized. In order to select an initial point for curve fitting, let us recall that, as shown in Subsection \ref{subsec:DMA_model} and demonstrated in Figs. \ref{fig:Lorentz1Mag}-\ref{fig:Lorentz1Phase}, the Lorentzian function describing the frequency response of the $l$th element of microstrip $i$ \eqref{eqn:Form_of_Weight} represents one of the following three families of frequency selective profiles:
	\begin{enumerate}
		\item {\em Monotonically decreasing magnitude} - achieved by setting the resonance frequency $\Omega_{i,l}^R$ to be smaller than the lowest frequency in the band of interest. In particular, setting $\Omega_{i,l}^R = 0$ and $\chi_{i,l}=0$ yields amplitude-only frequency flat weights.
		\item {\em Monotonically increasing magnitude} - such profiles are obtained by setting $\Omega_{i,l}^R$ to be larger than the maximal frequency in the band of interest, where the slope is determined by how far $\Omega_{i,l}^R$ is from the band of interest.
		\item {\em  Unimodal profile} - when  $\Omega_{i,l}^R$  is within the considered frequency band, the element resembles a unimodal function, i.e., a bandpass filter, centered at $\Omega_{i,l}^R$.
	\end{enumerate} 
	In addition to the aforementioned profiles, which describe the spectral behavior of the magnitude of the elements frequency response, it is also noted that the phase of \eqref{eqn:Form_of_Weight} is always in the upper half of the complex plain, i.e., in the range $[0,\pi]$, as also observed in Fig. \ref{fig:Lorentz1Phase}.
	
	The feasible spectral profiles indicate which types of unconstrained weights are best captured by \ac{dma} elements. In particular, if for some $(i,l)\in \calN_d\times \calN_e$, the unconstrained weights $\{\alpha_{m,i}(\hat{\qq}_{m,i})_l\}_{n \in \calM}$ exhibit a decreasing magnitude, then it is likely they can be well-approximated by the output of non-linear least squares solver starting from a resonance frequency smaller than the lower edge of the band of interest, e.g., $\Omega_{i,l}^R = f_c  - f_s/2 - \Delta$, for some fixed $\Delta > 0$. Similarly, when the behavior of $\{\alpha_{m,i}(\hat{\qq}_{m,i})_l\}_{n \in \calM}$ resembles a monotonic increase in magnitude, they should be accurately approached when curve fitting starting from $\Omega_{i,l}^R = f_c  + f_s/2 + \Delta$. Finally, if the amplitudes of $\{\alpha_{m,i}(\hat{\qq}_{m,i})_l\}_{n \in \calM}$ are a unimodal curve whose peak, located in index $\tilde{m} \in \calM$,  lies in the upper half of the complex plain, then using $\Omega_{i,l}^R = \Omega_{\tilde{m}}$ as a starting point is expected to yield a close approximation of the unconstrained weights. This understanding of the spectral behavior of \ac{dma} elements can be used by choosing those three suggested initial points for gradient search optimization, and taking the setting which achieves the minimal objective \eqref{eqn:ProbProjFreq}. This approach is summarized as Algorithm \ref{A3}. As in the frequency flat case, the resulting weights $\bar{\qQ}$ are used to determine the \ac{adc} support $\gamma$ and the digital filter via \eqref{eqn:Support} and \eqref{Mat_min_mmse}, respectively.
	
	\begin{algorithm}
		\SetKwInOut{Input}{Input}
		\caption{Frequency-selective setting\label{A3}}
\ifsingle
\else
		\small
\fi
		\Input{Unconstrained weights $\{\hat{\qq}_{m,i}\}$, $m\in\calM$, $i \in \calN_d$; \newline 
			alternating iterations limit ${\rm iter}_{\max}$.}
		\KwData{$\alpha_{m,i} = 1$ for all $(m,i)\!\in\!\calM \!\times\!\calN_d$. \\
			$F_{i,l} = 1$ for all $(i,l)\!\in\!\calN_d 	\!\times\!\calN_e$.}
		\For{$i=1,2,\ldots,N_d$}{ 
			\For{${\rm iter}=1,2,\ldots,{\rm iter}_{\max}$}{
				\For{$l=1,2,\ldots,N_e$}{ 
					Set $(\Omega_{i,l}^{R,(1)}, \chi_{i,l}^{(1)})$ using a non-linear least-squares solver to fit \eqref{eqn:ProbProjFreq} starting from  $\Omega_{i,l}^R = f_c  - f_s/2 - \Delta$\;
					Set $(\Omega_{i,l}^{R,(2)}, \chi_{i,l}^{(2)})$ using a non-linear least-squares solver to fit \eqref{eqn:ProbProjFreq} starting from  $\Omega_{i,l}^R = f_c + f_s/2 + \Delta$\;
					Set $(\Omega_{i,l}^{R,(3)}, \chi_{i,l}^{(3)})$ using a non-linear least-squares solver to fit \eqref{eqn:ProbProjFreq} starting from  $\Omega_{i,l}^R = \Omega_{\tilde{m}}$\;
					Set $(\hat{\Omega}_{i,l}^{R}, \hat{\chi}_{i,l})$ as  the pair from $\{\Omega_{i,l}^{R,(k)}, \chi_{i.l}^{(k)}\}$ which minimizes \eqref{eqn:ProbProjFreq}\;
					Set $\hat{F}_{i,l}$ using  Lemma \ref{lem:SetGain}\;				
				}
				Set $\{\alpha_{m,i}\}_{m=1}^{M} $ using Lemma \ref{lem:Sol_Projection2}\;			
			}
		}
		\KwOut{$({\qQ}_m)_{i,l}=q_{i,l}(\omega_m)$ via \eqref{eqn:Form_of_Weight} for all $(m,i,l) \in \calM \times \calN_d \times \calN_e$.}
	\end{algorithm}

	\vspace{-0.2cm}
	\subsection{Discussion}
	\label{subsec:Discussion}
	\vspace{-0.1cm}
	\ac{dma}-based receivers, as follows from our model in Subsection \ref{subsec:DMA_model}, implement a type of hybrid beamforming. In particular, the received signal undergoes some processing which reduces its dimensionality prior to being converted into a digital representation.  Such hybrid architectures are commonly used in the massive \ac{mimo} literature, often as method to reduce the number of RF chains \cite{mendez2016hybrid, ioushua2019family,sohrabi2017hybrid} but also to facilitate recovery under bit constraints \cite{shlezinger2018asymptotic}. 
	Conventional hybrid receivers utilize an additional dedicated hardware to implement the analog combining, typically consisting of phase shifter networks, i.e., an interconnection of phase shifters and adders.
	One of the benefits of using \acp{dma} over conventional hybrid architectures, also noted in \cite{shlezinger2019dynamic}, is that the controllable analog combining in \acp{dma} is a natural byproduct of the antenna structure, and does not require additional dedicated hardware. Another advantage stems from the fact that conventional hybrid systems, such as phase shifter networks, are typically frequency flat, namely, the same analog mapping is implemented for all the spectral components of the input signal \cite{sohrabi2017hybrid}. The inherent adjustable frequency selectivity of the \ac{dma} elements implies that they can be tuned to apply a frequency varying analog combining, which corresponds to one of the three profiles discussed in the previous subsection. This property is expected to improve the achievable performance in wideband frequency selective setups, as numerically demonstrated in Section \ref{sec:Sims}.
	
	Furthermore, the proposed design methods are based on task-based quantization schemes,  exploiting the block diagonal structure of $\bar{\qQ}$ to yield a set of optimization problems which can be solved in a greedy fashion. This structure of $\bar{\qQ}$ stems from the \ac{dma} architecture, which consists of several microstrips each feeding a different \ac{adc}. As the \acp{dma} operation is modeled as a form of analog beamforming, the proposed approach can also be utilized for conventional hybrid structures in which the  analog combiner obeys a block diagonal model, i.e., partially connected networks \cite{mendez2016hybrid, roth2018comparison}. In these conventional architectures, the filter $\qH(\omega)$, modeling the propagation inside the microstrips, is replaced with the identity mapping. 
	Consequently, our approach, which is based on formulating the signal recovery problem as a task-based quantization system, is expected to also facilitate the design of standard hybrid receivers. As we focus here on \ac{dma}-based receivers, we leave the analysis of the application of our methods for tuning conventional hybrid systems for future work.
	
	The spectral profile of the \ac{dma} elements is exploited in Algorithm \ref{A3} to achieve various types of frequency selective analog combining. In particular, the Lorentzian form \eqref{eqn:Form_of_Weight} allows to set each element to approach a controllable bandpass filter or an approximately spectrally-linear gain. While these profiles accommodate a broad family of spectral shapes, more complex functions, such as multi-modal frequency responses, may not be closely approximated by \acp{dma}. Furthermore, in Algorithm \ref{A3} we allowed to set the resonance frequencies $\{\Omega_{i,l}^R\}$ and the damping factors $\{\chi_{i,l}\}$ to be any non-negative values. However, setting the elements to have a large quality factor $\frac{\Omega_{i,l}^R}{\chi_{i,l}}$ may be difficult in practice, and this ratio is commonly restricted to be in the order of several tens \cite{Hunt-2013Science}. These additional considerations can also be accounted for in the optimization of the frequency response parameters of the elements. For example, in our numerical study in Section \ref{sec:Sims} we restrict the quality factor to a set of feasible values. 

	The algorithms detailed in the previous subsections provide methods for configuring the \ac{dma} weights along with the \acp{adc} and the digital processing to accurately 
	recover \ac{ofdm} signals, facilitating their decoding by the \ac{bs}. This procedure requires the receiver to know the multipath channel $\qG[\tau]$, the noise covariance $\qC_W$, and the \ac{dma} frequency selectivity profile $\qH(\omega)$. While the latter can be obtained from the physics of the metasurface, the channel parameters must be estimated, which may be a challenging task in the presence of quantized channel outputs \cite{shlezinger2018asymptotic,wang2019reliable,mo2017channel}. When channel estimation is carried out in a time division duplexing manner, the dynamic nature of \acp{dma} can be exploited to assign different configurations during channel estimation and signal recovery. However, we leave this study for future investigation.

	\section{Numerical Evaluations}
	\label{sec:Sims}
	\vspace{-0.1cm}
	In this section, we numerically evaluate the performance of bit-constrained MIMO-OFDM systems in which the \ac{bs} is equipped with a \ac{dma} configured using the method detailed in Section \ref{sec:DMA}, demonstrating the performance gains achieved by exploiting the frequency selectivity of \acp{dma} and by treating the receiver operation as a task-based quantization setup. 
	We consider a single-cell uplink MU-MIMO setup in which a \ac{bs} serves $K=8$ users. The users simultaneously transmit \ac{ofdm} signals with central frequency $f_c = 1.9$ GHz, each comprised of $M=16$ subcarriers with carrier spacing of $20$ MHz. 
	The data symbols are independently drawn from a QPSK constellation.

	The wireless channel is generated based on the correlated Gaussian model for rich scattering environments \cite{xiao2004discrete} with $L_{G}=4$ taps. The resulting discrete-time channel is given by $\qG[\tau]=\sigma^2_{\qG}[\tau]\qSigma_R^{\frac{1}{2}}\qG_R[\tau]$,
	where $\{\sigma^2_{\qG}[\tau]\}$ represents a temporal exponentially decaying profile, i.e., $\sigma^2_{\qG}[\tau]=e^{-\tau}$;
	$\{\qG_R[\tau]\}$ are $N	\times K$ Rayleigh fading matrices;
	and $\qSigma_R$ is the correlation matrix of the antenna array, induced by sub-wavelength element spacing. In particular, we set $\qSigma_R=\qI_{N_d}\otimes\qSigma_C$, modelling the case where the microstrips are sufficiently spaced such that elements of different micropstrips are not correlated, where $\qSigma_C \in \calC^{N_e \times N_e}$ represents the spatial correlation among the elements of the same microstrip with $0.2$ wavelength spacing. Following Jakes' model \cite{jakes1994microwave}, we use $(\qSigma_C)_{i,l}=J_0(0.4\pi|i-l|)$, where $J_0(\cdot)$ is the zero-order Bessel function of the first type and $i,l\in\calN_e$. 
	Due to the coupling between elements, the noise is spatially correlated, and we choose $\qC_W = \sigma_z^2 \qSigma_R$, with $\sigma_z^2 > 0$.

	We consider a \ac{dma} comprised of a total $N=100$ elements, divided into $N_d = 10$ microstrips with $N_e = 10$ elements in each.
	The propagation inside the microstrip, modeled via the diagonal matrix ${\qH}(\omega)$, is set to $({\qH}(\omega))_{l,l}=e^{-\alpha l -j\beta(\omega)l}$, where $\alpha = 0.006$ $[{\text m}^{-1}]$ and $\beta (\omega) = 1.592\cdot \omega$ $[{\text m}^{-1}]$. This setting represents microstrips with 50 ohm
	characteristic impedance made of Duroid 5880 operating at $1.9$ GHz with element spacing of
	$0.2$ wavelength (assuming free space wavelength) \cite[Ch. 3.8]{pozar2009microwave}.  
	
	In the following we numerically evaluate the performance in terms of signal recovery \ac{mse} and uncoded \ac{ber} of the following receivers:
	\begin{enumerate}[label={\em R\arabic*}]
		\item \label{itm:Unconst} A \ac{dma} with unconstrained weights designed using Algorithm \ref{A0}. 
		\item \label{itm:FreqSel} A \ac{dma} whose elements obey the Lorentzian model \eqref{eqn:FreqSel}, designed to approximate the unconstrained \ac{dma} setting \ref{itm:Unconst} using Algorithm \ref{A3}. To guarantee that the elements are configured with a feasible quality factor, we fix the ratio $\frac{\Omega_{i,l}^R}{\chi_{i,l}}$ to be in the discrete set $[0.1, 5, 30]$, and carry out the non-linear least-squared curve fitting in Algorithm \ref{A3}, implemented using the Levenberg-Marquardt method \cite{marquardt1963algorithm}, with respect to the resonance frequency $\Omega_{i,l}^R$ for each $(i,l) \in \calN_d \times \calN_e$. 
		\item \label{itm:AmpOnly} A \ac{dma} whose elements are restricted to be frequency flat, following the approximation used in \cite{shlezinger2019dynamic}. In particular, we consider amplitude only weights, for which the response is selected in the range $[0.001, 1]$ to approximate the unconstrained \ac{dma} \ref{itm:Unconst} using Algorithm~\ref{A1}. 
		\item \label{itm:RefPhaseShifter} A hybrid receiver with a partially-connected phase shifter network designed using the method of \cite{roth2018comparison}, operating with fixed \acp{adc} of support $\gamma = 100$. The support value was selected based on a set of numerical tests where it was shown to guarantee \ac{adc} overloading probability of roughly $1\%$.  This receiver represents previously proposed hybrid architectures for bit-constrained scenarios which are not derived using the task-based quantization framework, as we do in our work. 
		\item \label{itm:LMMSE} The linear \ac{mmse} estimator for recovering the symbols from the channel output {\em without quantization constraints}. This receiver represents a lower bound on the achievable performance of the bit-constrained receivers \ref{itm:Unconst}-\ref{itm:RefPhaseShifter}.
	\end{enumerate}
	For the bit-constrained \acp{bs} \ref{itm:Unconst}-\ref{itm:RefPhaseShifter}, we consider an overall  budget of up to $b_{\rm overall}$ bits, divided equally among the \acp{adc} such that each \ac{adc} operates with a resolution of $b = \lfloor2^{b_{\rm overall}/(2\cdot N_d)} \rfloor$ decision regions. For the task-based quantizers \ref{itm:Unconst}-\ref{itm:AmpOnly} we used $\eta = 2$. 
	The output of the \acp{adc} is processed using the digital filter of Lemma \ref{lem:taskBased}. The results are computed by averaging over $10^3$ \ac{mimo}-\ac{ofdm} symbols.

\begin{figure}
		\centering
		\includegraphics[width=\figWidth]{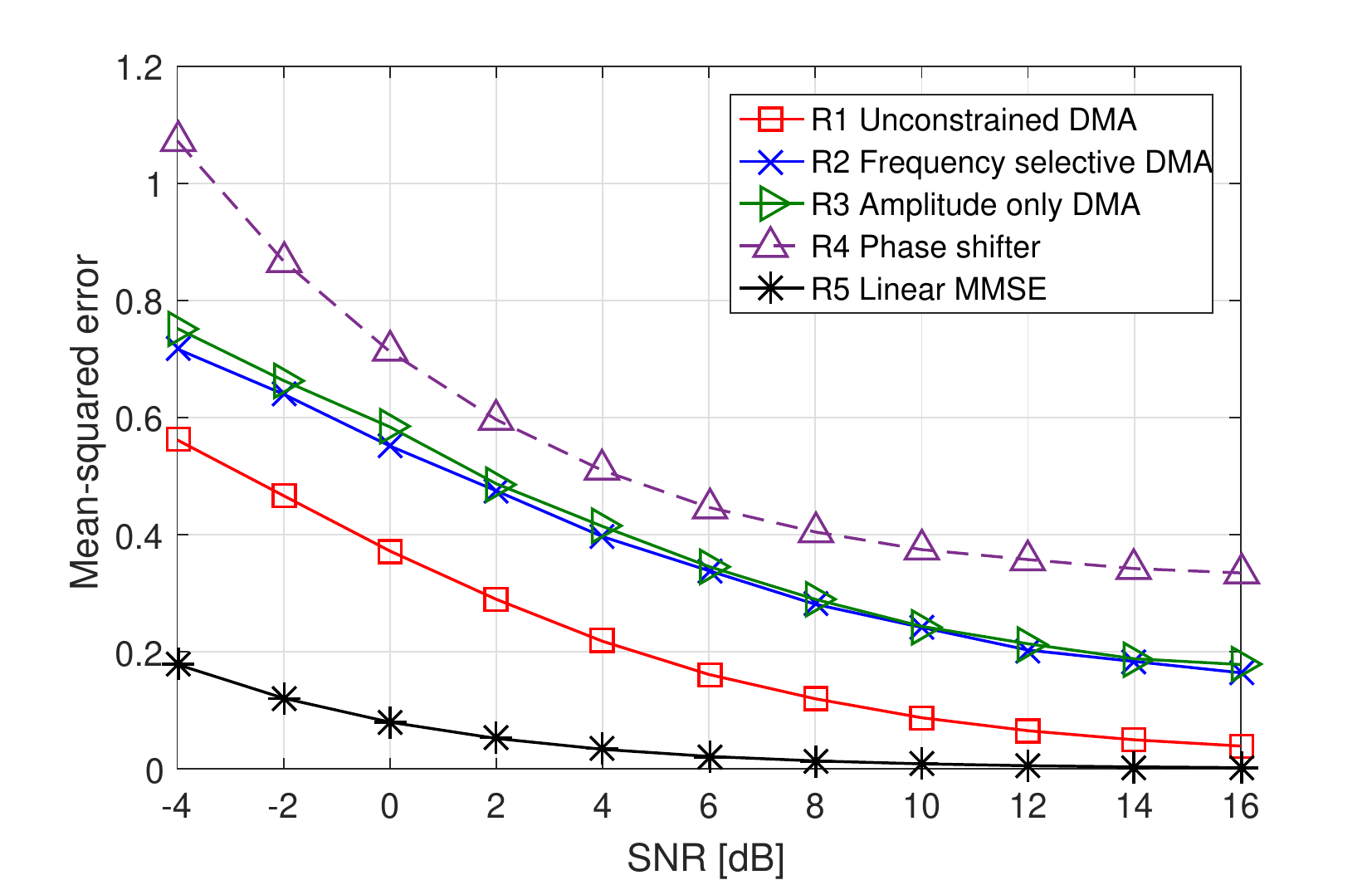}
		\vspace{-0.2cm}
		\caption{\ac{mse} versus \ac{snr}, $80$ overall bits.\label{F1}}
	\end{figure}

	We begin by evaluating the signal recovery \ac{mse} of the considered receiver structures versus the \ac{snr} in the range of $[-4,16]$ dB, for an overall bit budget of $b_{\rm overall} = 80$ bits. 
	 The results are depicted in Fig. \ref{F1}. Observing Fig. \ref{F1}, we first note that the unconstrained \ac{dma} \ref{itm:Unconst}, designed based on the task-based quantization framework of \cite{shlezinger2018hardware}, achieves the most accurate recovery among the bit-constrained receivers, indicating that the proposed greedy method in Algorithm \ref{A0} yields useful configurations. The frequency selective \ac{dma} \ref{itm:FreqSel} designed using Algorithm \ref{A3} outperforms the frequency flat hybrid receivers \ref{itm:AmpOnly}-\ref{itm:RefPhaseShifter}, with substantial gains over the previously proposed phase shifter network of \cite{roth2018comparison}, and more minor gains over the amplitude-only \ac{dma} 
	 designed using our proposed Algorithm \ref{A1}. In particular, \ref{itm:FreqSel} achieves an \ac{mse} of $0.4$ at \ac{snr} of $4$ dB, while the previously proposed \ref{itm:RefPhaseShifter} requires \ac{snr} of at least $8$ dB to achieve the same \ac{mse}, i.e., an \ac{snr} gain of $4$ dB. The corresponding \ac{snr} gain of the frequency selective \ref{itm:FreqSel} over the frequency flat \ref{itm:AmpOnly}, designed using our Algorithm \ref{A1}, is $0.5$ dB. In addition to performance gains of the \ac{dma}-based \acp{bs} over the phase shifter receiver \ref{itm:RefPhaseShifter}, the latter requires additional dedicated hardware for combining the observed signals, while \acp{dma} implements this analog combining as a natural byproduct of their antenna architecture. These results demonstrate the benefits of exploiting the physical characteristics of \acp{dma} using a task-based quantization framework, in which the analog combining is jointly optimized with the \ac{adc} support and the digital processing. 
	
	\begin{figure}
		\centering
		\includegraphics[width=\figWidth]{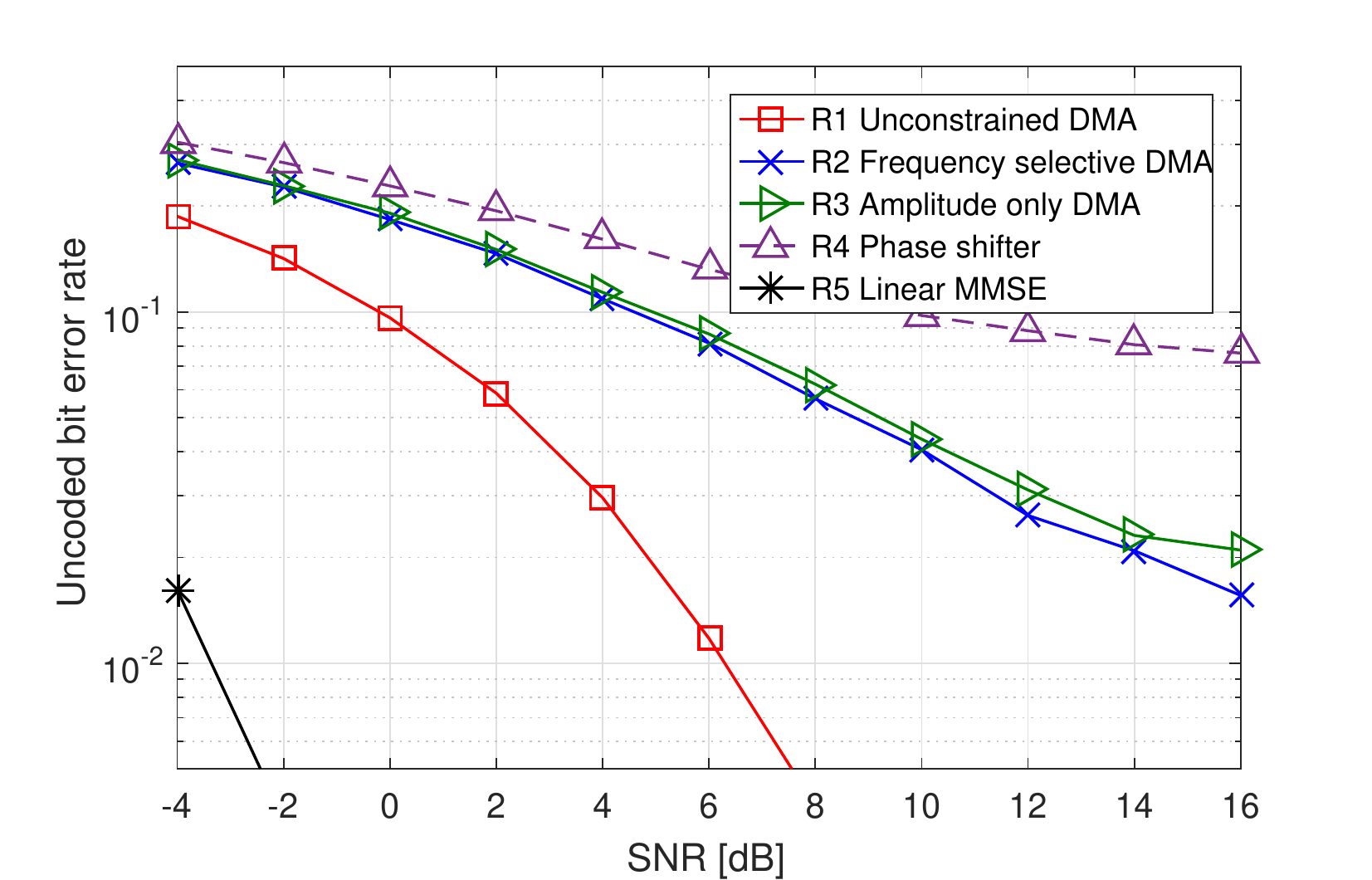}
		\vspace{-0.2cm}
		\caption{Uncoded \ac{ber} versus \ac{snr}, $80$ overall bits.\label{F2}}
	\end{figure}
	
	To evaluate how the \ac{mse} improvement of our proposed approach is translated into gains in  \ac{ber}, we depict in Fig. \ref{F2} the uncoded \ac{ber} of the above receivers versus \ac{snr} for $b_{\rm overall} = 80$ bits. The results in Fig. \ref{F2} demonstrate that the gains in \ac{mse} observed in Fig. \ref{F1} are translated in uncoded \ac{ber} improvement of a lesser magnitude. In particular, \ref{itm:FreqSel} achieves an uncoded \ac{ber} of $8\cdot10^{-2}$ at \ac{snr} of $6$ dB, while \ref{itm:AmpOnly} and \ref{itm:RefPhaseShifter} achieve the same \ac{ber} value for \acp{snr} of roughly $6.5$ dB and $14$ dB, respectively, namely, an \ac{snr} gain of $0.5$ dB and $7.5$ dB, respectively. 
	
	\begin{figure}
	\centering
	\includegraphics[width=\figWidth]{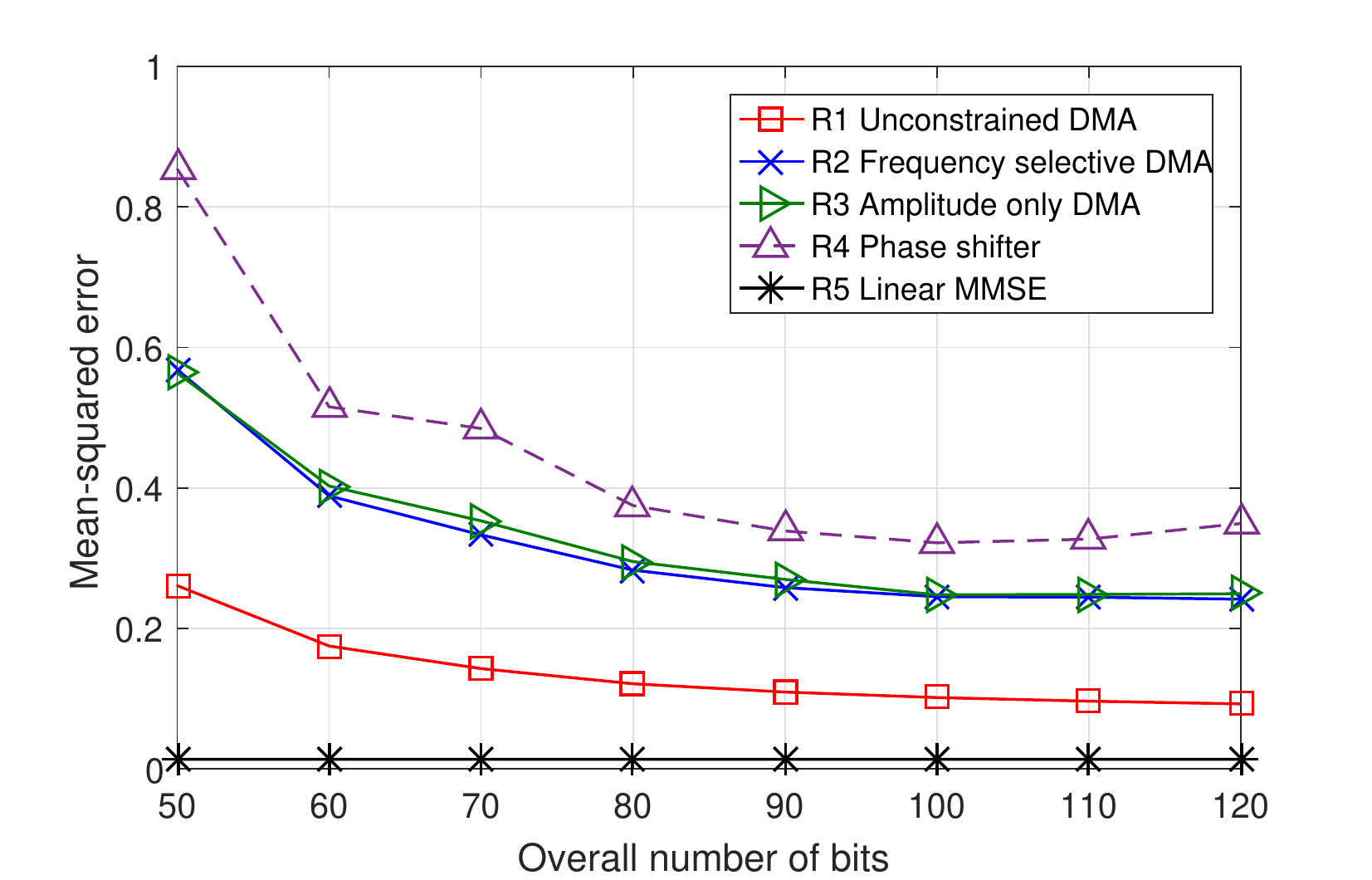}
	\vspace{-0.2cm}
	\caption{\ac{mse} versus bits, \ac{snr} of $8$ dB.\label{F3}}
\end{figure}

\begin{figure}
	\centering
	\includegraphics[width=\figWidth]{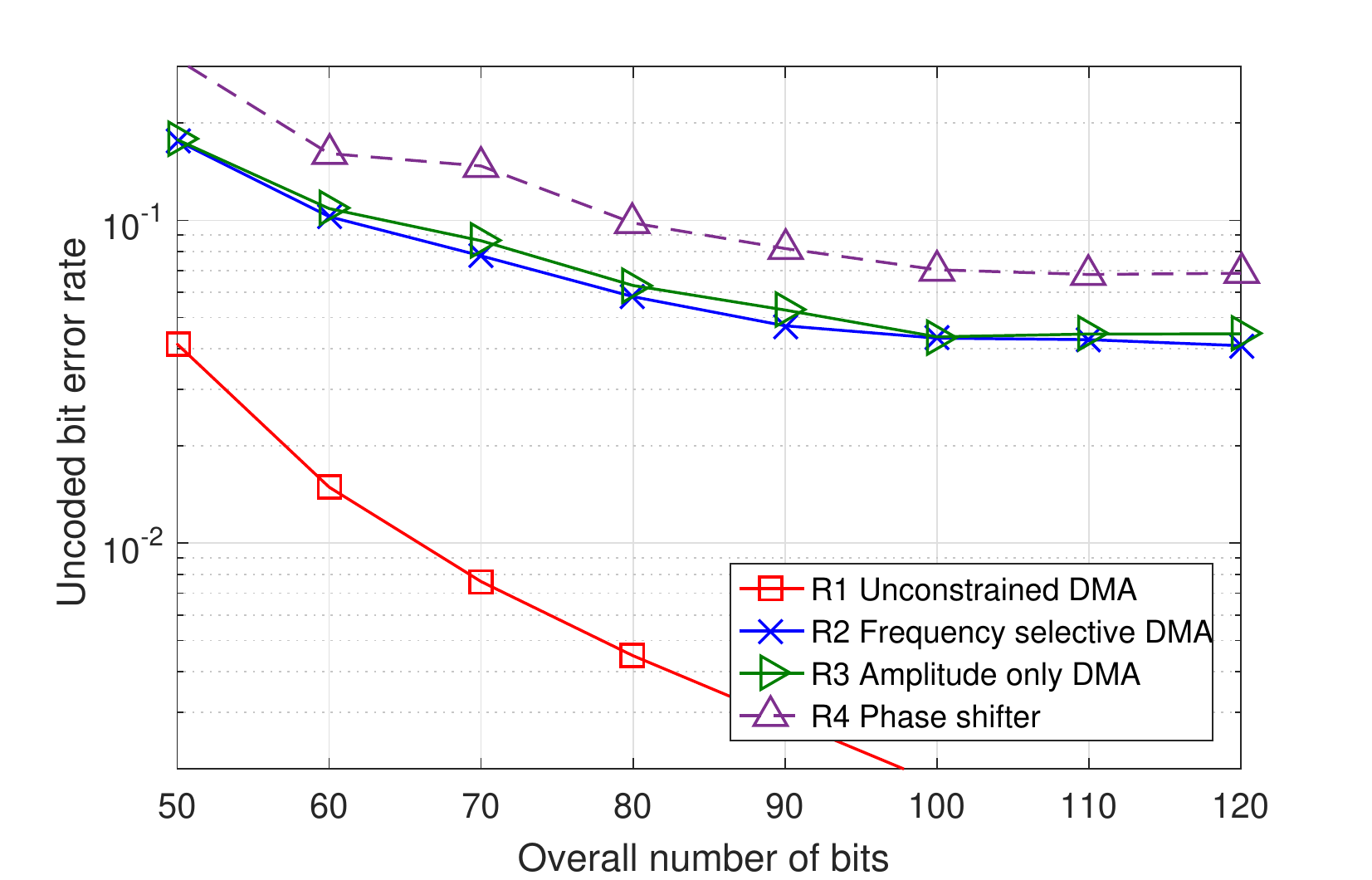}
	\vspace{-0.2cm}
	\caption{Uncoded \ac{ber} versus bits, \ac{snr} of $8$ dB.\label{F4}}
\end{figure}
	
	Finally, we evaluate how the performance of the proposed \acp{dma} configurations scales with respect to the overall bit budget, compared to phase shifter based hybrid receivers as well as the unquantized linear \ac{mmse} estimator. To that aim, we depict in Figs. \ref{F3}-\ref{F4} the resulting signal recovery \ac{mse} and uncoded \ac{ber}, respectively, of the considered receivers versus the bit budget $b_{\rm overall} \in \{50,120\}$ for an \ac{snr} of $8$ dB. The \ac{ber} achieved by the linear \ac{mmse} estimator operating without quantization constraints is significantly lower than that achieved by the bit-constrained receiver, and is thus not included in Fig. \ref{F4}. Note that a fully-digital receiver, i.e., one in which each antenna element feeds a dedicated \ac{adc} as commonly assumed in  \ac{mimo} communications receivers \cite{wang2019reliable,jacobsson2017throughput,choi2016near,mo2017channel}, cannot be applied with less than $2N = 200$ bits. Consequently, all the considered values of $b_{\rm overall}$ correspond to \acp{bs} operating under strict bit constraints. Observing Figs. \ref{F3}-\ref{F4}, we note that the \ac{mse} and \ac{ber} gains of the proposed designs, observed in Figs. \ref{F1}-\ref{F2} for receivers operating utilizing $80$ bits for representing their channel output, hold for all considered values of $b_{\rm overall}$.

	 To summarize, \acp{dma}, which typically use less power and cost less than standard antenna arrays, can be utilized to implement a configurable frequency selective hybrid architecture. The results presented in this section demonstrate that by using our proposed methods, one can design a high-performance \ac{dma}-based receiver which is particularly suitable for bit-constrained MU-MIMO-OFDM setups, achieving notable performance gains over conventional phase shifter based receivers designed using previously proposed methods.

	\section{Conclusions}
	\label{sec:Conclusions}
	In this work we studied the application of DMAs, which realize low cost and power efficient  configurable antenna arrays, for bit-constrained MU-MIMO-OFDM systems. We formulated a model for the received quantized DMA outputs which accounts for the adaptable frequency selective profile of the metamaterial elements, and showed that the OFDM recovery problem can be expressed as a task-based quantization setup. Next, we derived an iterative algorithm for setting the DMA weights, ignoring their structure constraints, to minimize the MSE in recovering the transmitted signal, by sequentially adapting each microstrip. Then, we proposed methods for approximating the unconstrained \ac{dma} using configurations which capture the physical properties of metasurfaces, considering both narrowband approximations of their operation as well as the more general Lorentzain-type wideband model.  Our numerical results demonstrate the gains of utilizing task-based quantization to determine the pre-quantization combining carried out in the \ac{dma} along with the \ac{adc} support and the digital processing. In particular, it is shown that by properly exploiting the physical characteristics of DMAs, \ac{ofdm} symbol detection is significantly facilitated when operating   under strict bit constraints. 
	
	\ifproofs	
	\begin{appendix}
		\numberwithin{proposition}{subsection} 
		\numberwithin{lemma}{subsection} 
		\numberwithin{corollary}{subsection} 
		\numberwithin{remark}{subsection} 
		\numberwithin{equation}{subsection}	
		%
		\vspace{-0.2cm}

		\subsection{Proof of Proposition \ref{pro:BlkDiagStructure}}
		\label{app:Proof1}
		It follows directly from the definition of $\qQ$ that $\qQ^H\qQ=\mathop{\rm blkdiag}(\bar{\qq}_1\bar{\qq}^{H}_1,\bar{\qq}_2\bar{\qq}^{H}_2,\cdots,\bar{\qq}_{N_d}\bar{\qq}^{H}_{N_d})$.
		We define the following matrices to facilitate subsequent proof:
		\[\begin{aligned}
		{\qE_j} &= [ {\underbrace {{\qO_{{N_e}}}\, \cdots \;{\qO_{{N_e}}}}_{j-1}\;{\qI_{{N_e}}}\;\underbrace {{\qO_{{N_e}}}\, \cdots \;{\qO_{{N_e}}}}_{N_d-j}} ]^{T},\\
		\qPhi_j &= [ {\underbrace {{\qO_{{N_e}}}\, \cdots \;{\qO_{{N_e}}}}_{j-1}\;(\bar{\qq}_j\bar{\qq}^{H}_j)^{T}\;\underbrace {{\qO_{{N_e}}}\, \cdots \;{\qO_{{N_e}}}}_{N_d-j}} ]^{T}.
		\end{aligned}
		\]
		Clearly, $\qPhi_j=\qE_j\bar{\qq}_j\bar{\qq}^{H}_j\qE^H_j$. 
		Then the matrix $\qQ^H\qQ$ can be derived as
		\begin{equation}\label{eqn:ProofAid2}
		\qQ^H\qQ=\sum_{j=1}^{N_d}\qPhi_j=\sum_{j=1}^{N_d}\qE_j\bar{\qq}_j\bar{\qq}^{H}_j\qE^H_j.
		\end{equation}
		Substituting \eqref{eqn:ProofAid2} into \eqref{min_mmse} along with the block-diagonal expressions for $\bar{\qG}$ and $\qSigma$, and the fact that the inverse of a block-diagonal matrix is block-diagonal with the inverse submatrices \cite[Ch. 3.7]{meyer2000matrix}, yields \eqref{min_mmse_simp}, proving the proposition. 
		$\qed$
	\end{appendix}
	
	\fi	
	\bibliographystyle{IEEEtran}
	\bibliography{IEEEabrv,DMABib}

\begin{thebibliography}{10}
\providecommand{\url}[1]{#1}
\csname url@samestyle\endcsname
\providecommand{\newblock}{\relax}
\providecommand{\bibinfo}[2]{#2}
\providecommand{\BIBentrySTDinterwordspacing}{\spaceskip=0pt\relax}
\providecommand{\BIBentryALTinterwordstretchfactor}{4}
\providecommand{\BIBentryALTinterwordspacing}{\spaceskip=\fontdimen2\font plus
\BIBentryALTinterwordstretchfactor\fontdimen3\font minus
  \fontdimen4\font\relax}
\providecommand{\BIBforeignlanguage}[2]{{%
\expandafter\ifx\csname l@#1\endcsname\relax
\typeout{** WARNING: IEEEtran.bst: No hyphenation pattern has been}%
\typeout{** loaded for the language `#1'. Using the pattern for}%
\typeout{** the default language instead.}%
\else
\language=\csname l@#1\endcsname
\fi
#2}}
\providecommand{\BIBdecl}{\relax}
\BIBdecl

\bibitem{Marzetta-2010TWC}
T.~L. Marzetta, ``Noncooperative cellular wireless with unlimited numbers of
  base station antennas,'' \emph{{IEEE} Trans. Wireless Commun.}, vol.~9,
  no.~11, pp. 3590--3600, Nov. 2010.

\bibitem{jiang2007multiuser}
M.~Jiang and L.~Hanzo, ``Multiuser {MIMO-OFDM} for next-generation wireless
  systems,'' \emph{Proc. {IEEE}}, vol.~95, no.~7, pp. 1430--1469, 2007.

\bibitem{Eldar2015}
Y.~C. Eldar, \emph{Sampling theory: Beyond bandlimited systems}.\hskip 1em plus
  0.5em minus 0.4em\relax Cambridge University Press, 2015.

\bibitem{walden1999analog}
R.~H. Walden, ``Analog-to-digital converter survey and analysis,'' \emph{{IEEE}
  J. Sel. Areas Commun.}, vol.~17, no.~4, pp. 539--550, 1999.

\bibitem{Andrews-2014JSAC}
J.~G. Andrews, S.~Buzzi, W.~Choi, S.~V. Hanly, A.~Lozano, A.~C. Soong, and
  J.~C. Zhang, ``What will {5G} be?'' \emph{{IEEE} J. Sel. Areas Commun.},
  vol.~32, no.~6, pp. 1065--1082, 2014.

\bibitem{shlezinger2018hardware}
N.~Shlezinger, Y.~C. Eldar, and M.~R. Rodrigues, ``Hardware-limited task-based
  quantization,'' \emph{{IEEE} Trans. Signal Process.}, vol.~67, no.~20, pp.
  5223--5238, 2019.

\bibitem{salamatian2019task}
S.~Salamatian, N.~Shlezinger, Y.~C. Eldar, and M.~M{\'e}dard, ``Task-based
  quantization for recovering quadratic functions using principal inertia
  components,'' in \emph{Proc. IEEE ISIT}, 2019.

\bibitem{shlezinger2019deep}
N.~Shlezinger and Y.~C. Eldar, ``Deep task-based quantization,'' \emph{arXiv
  preprint arXiv:1908.06845}, 2019.

\bibitem{zhang2005variable}
X.~Zhang, A.~F. Molisch, and S.-Y. Kung, ``Variable-phase-shift-based
  {RF}-baseband codesign for {MIMO} antenna selection,'' \emph{{IEEE} Trans.
  Signal Process.}, vol.~53, no.~11, p. 4091, 2005.

\bibitem{mendez2016hybrid}
R.~M{\'e}ndez-Rial, C.~Rusu, N.~Gonz{\'a}lez-Prelcic, A.~Alkhateeb, and R.~W.
  Heath, ``Hybrid {MIMO} architectures for millimeter wave communications:
  Phase shifters or switches?'' \emph{IEEE Access}, vol.~4, pp. 247--267, 2016.

\bibitem{ioushua2019family}
S.~S. Ioushua and Y.~C. Eldar, ``A family of hybrid analog--digital beamforming
  methods for massive {MIMO} systems,'' \emph{{IEEE} Trans. Signal Process.},
  vol.~67, no.~12, pp. 3243--3257, 2019.

\bibitem{sohrabi2017hybrid}
F.~Sohrabi and W.~Yu, ``Hybrid analog and digital beamforming for mmwave {OFDM}
  large-scale antenna arrays,'' \emph{{IEEE} J. Sel. Areas Commun.}, vol.~35,
  no.~7, pp. 1432--1443, 2017.

\bibitem{shlezinger2018asymptotic}
N.~Shlezinger, Y.~C. Eldar, and M.~R. Rodrigues, ``Asymptotic task-based
  quantization with application to massive {MIMO},'' \emph{{IEEE} Trans. Signal
  Process.}, vol.~67, no.~15, pp. 3995--4012, 2019.

\bibitem{gong2019rf}
T.~Gong, N.~Shlezinger, S.~S. Ioushua, M.~Namer, Z.~Yang, and Y.~C. Eldar,
  ``{RF} chain reduction for {MIMO} systems: A hardware prototype,''
  \emph{arXiv preprint arXiv:1905.05315}, 2019.

\bibitem{DSmith-2017PRA}
D.~R. Smith, O.~Yurduseven, L.~Pulido-Mancera, P.~Bowen, and N.~B. Kundtz,
  ``Analysis of a waveguide-fed metasurface antenna,'' \emph{Phys. Rev.
  Applied}, vol.~8, no.~5, Nov. 2017.

\bibitem{smith2017analysis}
D.~R. Smith, V.~R. Gowda, O.~Yurduseven, S.~Larouche, G.~Lipworth, Y.~Urzhumov,
  and M.~S. Reynolds, ``An analysis of beamed wireless power transfer in the
  fresnel zone using a dynamic, metasurface aperture,'' \emph{Journal of
  Applied Physics}, vol. 121, no.~1, p. 014901, 2017.

\bibitem{Sleasman-2016JAWPL}
T.~Sleasman, M.~F. Imani, W.~Xu, J.~Hunt, T.~Driscoll, M.~S. Reynolds, and
  D.~R. Smith, ``Waveguide-fed tunable metamaterial element for dynamic
  apertures,'' \emph{{IEEE} Antennas Wireless Propag. Lett.}, vol.~15, pp.
  606--609, 2016.

\bibitem{Diebold-2018AO}
A.~V. Diebold, M.~F. Imani, T.~Sleasman, and D.~R. Smith, ``Phaseless
  computational ghost imaging at microwave frequencies using a dynamic
  metasurface aperture,'' \emph{Appl Opt.}, vol.~57, no.~9, pp. 2142--2149,
  2018.

\bibitem{huang2019reconfigurable}
C.~Huang, A.~Zappone, G.~C. Alexandropoulos, M.~Debbah, and C.~Yuen,
  ``Reconfigurable intelligent surfaces for energy efficiency in wireless
  communication,'' \emph{{IEEE} Trans. Wireless Commun.}, vol.~18, no.~8, pp.
  4157--4170, 2019.

\bibitem{di2019smart}
M.~Di~Renzo, M.~Debbah, D.-T. Phan-Huy, A.~Zappone, M.-S. Alouini, C.~Yuen,
  V.~Sciancalepore, G.~C. Alexandropoulos, J.~Hoydis, H.~Gacanin \emph{et~al.},
  ``Smart radio environments empowered by reconfigurable {AI} meta-surfaces: an
  idea whose time has come,'' \emph{EURASIP Journal on Wireless Communications
  and Networking}, vol. 2019, no.~1, pp. 1--20, 2019.

\bibitem{del2019optimally}
P.~del Hougne, M.~Fink, and G.~Lerosey, ``Optimally diverse communication
  channels in disordered environments with tuned randomness,'' \emph{Nature
  Electronics}, vol.~2, no.~1, p.~36, 2019.

\bibitem{tang2019wireless}
W.~Tang, X.~Li, J.~Y. Dai, S.~Jin, Y.~Zeng, Q.~Cheng, and T.~J. Cui, ``Wireless
  communications with programmable metasurface: Transceiver design and
  experimental results,'' \emph{China Communications}, vol.~16, no.~5, pp.
  46--61, 2019.

\bibitem{dai2019wireless}
J.~Y. Dai, W.~K. Tang, J.~Zhao, X.~Li, Q.~Cheng, J.~C. Ke, M.~Z. Chen, S.~Jin,
  and T.~J. Cui, ``Wireless communications through a simplified architecture
  based on time-domain digital coding metasurface,'' \emph{Advanced Materials
  Technologies}, vol.~4, no.~7, p. 1900044, 2019.

\bibitem{shlezinger2019dynamic}
N.~Shlezinger, O.~Dicker, Y.~C. Eldar, I.~Yoo, M.~F. Imani, and D.~R. Smith,
  ``Dynamic metasurface antennas for uplink massive {MIMO} systems,''
  \emph{{IEEE} Trans. Commun.}, vol.~67, no.~10, pp. 6829--6843, 2019.

\bibitem{wang2019dynamic}
H.~Wang, N.~Shlezinger, S.~Jin, Y.~C. Eldar, I.~Yoo, M.~F. Imani, and D.~R.
  Smith, ``Dynamic metasurface antennas based downlink massive {MIMO}
  systems,'' in \emph{Proc. IEEE SPAWC}, 2019, pp. 1--5.

\bibitem{DSmith-2018TCOM}
I.~Yoo, M.~F. Imani, T.~Sleasman, H.~D. Pfister, and D.~R. Smith, ``Enhancing
  capacity of spatial multiplexing systems using reconfigurable cavity-backed
  metasurface antennas in clustered {MIMO} channels,'' \emph{{IEEE} Trans.
  Commun.}, vol.~67, no.~2, pp. 1070--1084, Feb. 2018.

\bibitem{Johnson-2016TAP}
M.~C. Johnson, S.~L. Brunton, N.~B. Kundtz, and J.~N. Kutz, ``Sidelobe
  canceling for reconfigurable holographic metamaterial antennas,''
  \emph{{IEEE} Trans. Antennas Propag.}, vol.~63, no.~4, pp. 1881--1886, Apr.
  2015.

\bibitem{akyildiz2016realizing}
I.~F. Akyildiz and J.~M. Jornet, ``Realizing ultra-massive {MIMO} (1024$\times$
  1024) communication in the (0.06--10) terahertz band,'' \emph{Nano
  Communication Networks}, vol.~8, pp. 46--54, 2016.

\bibitem{roth2018comparison}
K.~Roth, H.~Pirzadeh, A.~L. Swindlehurst, and J.~A. Nossek, ``A comparison of
  hybrid beamforming and digital beamforming with low-resolution {ADCs} for
  multiple users and imperfect {CSI},'' \emph{{IEEE} J. Sel. Topics Signal
  Process.}, vol.~12, no.~3, pp. 484--498, June 2018.

\bibitem{mo2017channel}
J.~Mo, P.~Schniter, and R.~W. Heath, ``Channel estimation in broadband
  millimeter wave {MIMO} systems with few-bit {ADCs},'' \emph{{IEEE} Trans.
  Signal Process.}, vol.~66, no.~5, pp. 1141--1154, 2017.

\bibitem{jacobsson2017throughput}
S.~Jacobsson, G.~Durisi, M.~Coldrey, U.~Gustavsson, and C.~Studer, ``Throughput
  analysis of massive {MIMO} uplink with low-resolution {ADCs},'' \emph{{IEEE}
  Trans. Wireless Commun.}, vol.~16, no.~6, pp. 4038--4051, 2017.

\bibitem{choi2016near}
J.~Choi, J.~Mo, and R.~W. Heath, ``Near maximum-likelihood detector and channel
  estimator for uplink multiuser massive {MIMO} systems with one-bit {ADCs},''
  \emph{{IEEE} Trans. Commun.}, vol.~64, no.~5, pp. 2005--2018, 2016.

\bibitem{DRSmith-2004Science}
D.~R. Smith, J.~B. Pendry, and M.~C.~K. Wiltshire, ``Metamaterials and negative
  refractive index,'' \emph{Science}, vol. 305, no. 5685, p. 788–792, Aug.
  2004.

\bibitem{Smith-2012APM}
C.~L. {Holloway}, E.~F. {Kuester}, J.~A. {Gordon}, J.~{O'Hara}, J.~{Booth}, and
  D.~R. {Smith}, ``An overview of the theory and applications of metasurfaces:
  The two-dimensional equivalents of metamaterials,'' \emph{{IEEE} Antennas
  Propag. Mag.}, vol.~54, no.~2, pp. 10--35, Apr. 2012.

\bibitem{Hunt-2013Science}
J.~Hunt, T.~Driscoll, A.~Mrozack, G.~Lipworth, M.~Reynolds, D.~Brady, and D.~R.
  Smith, ``Metamaterial apertures for computational imaging,'' \emph{Science},
  vol. 339, no. 6117, pp. 310--313, 2013.

\bibitem{cover2012elements}
T.~M. Cover and J.~A. Thomas, \emph{Elements of information theory}.\hskip 1em
  plus 0.5em minus 0.4em\relax John Wiley \& Sons, 2012.

\bibitem{heath1999exploiting}
R.~W. Heath and G.~B. Giannakis, ``Exploiting input cyclostationarity for blind
  channel identification in {OFDM} systems,'' \emph{{IEEE} Trans. Signal
  Process.}, vol.~47, no.~3, pp. 848--856, 1999.

\bibitem{wei2010convergence}
S.~Wei, D.~L. Goeckel, and P.~A. Kelly, ``Convergence of the complex envelope
  of bandlimited {OFDM} signals,'' \emph{{IEEE} Trans. Inf. Theory}, vol.~56,
  no.~10, pp. 4893--4904, 2010.

\bibitem{hoydis2013massive}
J.~Hoydis, S.~Ten~Brink, and M.~Debbah, ``Massive mimo in the ul/dl of cellular
  networks: How many antennas do we need?'' \emph{{IEEE} J. Sel. Areas
  Commun.}, vol.~31, no.~2, pp. 160--171, 2013.

\bibitem{shlezinger2018spectral}
N.~Shlezinger and Y.~C. Eldar, ``On the spectral efficiency of noncooperative
  uplink massive {MIMO} systems,'' \emph{{IEEE} Trans. Commun.}, vol.~67,
  no.~3, pp. 1956--1971, 2019.

\bibitem{meyer2000matrix}
C.~D. Meyer, \emph{Matrix analysis and applied linear algebra}.\hskip 1em plus
  0.5em minus 0.4em\relax Siam, 2000.

\bibitem{Ghojogh-2019Arxiv}
B.~Ghojogh, F.~Karray, and M.~Crowley, ``Eigenvalue and generalized eigenvalue
  problems: Tutorial,'' \emph{arXiv preprint arXiv:1903.11240}, 2019.

\bibitem{marquardt1963algorithm}
D.~W. Marquardt, ``An algorithm for least-squares estimation of nonlinear
  parameters,'' \emph{Journal of the society for Industrial and Applied
  Mathematics}, vol.~11, no.~2, pp. 431--441, 1963.

\bibitem{wang2019reliable}
H.~{Wang}, W.~{Shih}, C.~{Wen}, and S.~{Jin}, ``Reliable {OFDM} receiver with
  ultra-low resolution {ADC},'' \emph{{IEEE} Trans. Commun.}, vol.~67, no.~5,
  pp. 3566--3579, May 2019.

\bibitem{xiao2004discrete}
C.~Xiao, J.~Wu, S.-Y. Leong, Y.~R. Zheng, and K.~B. Letaief, ``A discrete-time
  model for triply selective {MIMO} rayleigh fading channels,'' \emph{{IEEE}
  Trans. Wireless Commun.}, vol.~3, no.~5, pp. 1678--1688, 2004.

\bibitem{jakes1994microwave}
W.~C. Jakes and D.~C. Cox, \emph{Microwave mobile communications}.\hskip 1em
  plus 0.5em minus 0.4em\relax Wiley-IEEE Press, 1994.

\bibitem{pozar2009microwave}
D.~M. Pozar, \emph{Microwave engineering}.\hskip 1em plus 0.5em minus
  0.4em\relax John Wiley \& Sons, 2009.

\end{thebibliography}

\end{document}